\documentclass[12pt,tightenlines,eqsecnum,floats,aps,amsmath,amssymb,nofootinbib,prd]{revtex4}

\usepackage{setspace}
\usepackage{amsmath,amssymb,amsfonts,amsthm,mathrsfs}
\usepackage{graphicx}
\usepackage{enumerate}

\def\be{\begin{equation}}
\def\ee{\end{equation}}
\def\ba{\begin{eqnarray}}
\def\ea{\end{eqnarray}}
\def\bi{\begin{itemize}}
\def\ei{\end{itemize}}

\def\Z{\mathbb{Z}}
\def\reals{\mathbb{R}}

\def\qo{\mathring{q}}
\def\Do{\mathring{D}}

\def\eo{\mathring{e}}

\def\D{\mathcal{D}}
\def\L{\mathcal{L}}

\def\aut{\text{Aut}}
\def\diff{\text{Diff}}
\def\G{\mathcal{G}}
\def\tr{\text{Tr}}

\def\Eb{\bar{E}}
\def\bE{\beta_{\Eb}}
\def\bra{\langle}
\def\ket{\rangle}

\def\Haux{\mathcal{H}_{kin}}

\def\O{\mathcal{O}}

\def\Haut{\mathcal{H}_{\text{Aut}}}
\def\Vaut{\mathcal{V}_{\text{Aut}}}
\def\S{\mathcal{S}}
\def\Ph{\text{Ph}}

\def\nh{\hat{n}}
\def\id{\text{Id}}
\def\Lie{\text{Lie}}
\def\idtwo{\mathbf{1}}
\def\w{\omega}
\def\rk{\text{rank}}

\def\t{\tau}

\def\lqg{\text{\tiny LQG}}
\def\nn{\nonumber}

\def\P{\mathcal{P}}

\def\sym{\text{Sym}}
\def\qo{\mathring{q}}

\def\dim{\text{dim}}
\def\Vo{\mathring{V}}
\def\Vb{\overline{V}}
\def\Do{D}

\def\eo{\mathring{e}}

\def\sint{\textstyle{\int}}
\def\tg{\tilde{\gamma}}
\def\go{\mathring{\gamma}}
\def\tgo{\mathring{\tilde{\gamma}}}

\begin{document}
\title{The Koslowski- Sahlmann representation:
Gauge and diffeomorphism invariance
}
\author{Miguel Campiglia} \email{miguel@rri.res.in}
\author{Madhavan Varadarajan}\email{madhavan@rri.res.in}
 \affiliation{Raman Research Institute \\Bangalore-560 080, India}

\begin{abstract}

The  discrete spatial geometry underlying Loop Quantum Gravity (LQG) is
degenerate almost everywhere. This is at
apparent odds with the non- degeneracy of asymptotically flat metrics 
near spatial infinity.
Koslowski generalised 
the LQG representation so 
as to describe states labelled by smooth non- degenerate triad fields. 
His representation was further studied by Sahlmann with
a view to imposing gauge and spatial diffeomorphism invariance through group 
averaging methods. 
Motivated by the desire to model asymptotically  flat quantum geometry
by states with triad labels  which are non- degenerate at infinity but 
not necessarily so in the interior,
we initiate a generalisation of  
Sahlmann's considerations to triads of varying degeneracy. In doing so, we include
delicate phase contributions  to the averaging procedure which are crucial for the correct
implementation of the gauge and diffeomorphism constraints, and whose existence
can be traced to the background exponential functions recently constructed
by one of us. Our treatment emphasizes the role of symmetries of quantum states in the averaging procedure.
Semianalyticity, influential in  the proofs of the beautiful uniqueness results for LQG, plays a 
key role in our considerations. As a by product, we re- derive the group averaging map for standard LQG,
highlighting  the role of state symmetries and explicitly exhibiting the essential  uniqueness of its 
specification.

\end{abstract}
%\pacs{}
\maketitle

\section{Introduction} \label{sec1}
Loop Quantum Gravity (LQG) is an effort to construct a canonical quantization
of a classical 
Hamiltonian description of the gravitational field.
The phase space variables of this classical  Hamiltonian description are an 
$SU(2)$ connection and a conjugate electric (triad) field on a Cauchy slice
$\Sigma$. Most work in LQG is in the context of {\em compact} 
(without boundary)
Cauchy slices. In this context one of the key results of LQG is that its
underlying representation endows the quantum spatial geometry described
by the triad field with a fundamental {\em discreteness} \cite{discrete}.

It is of physical interest to  generalise LQG to the context
of asymptotically flat gravitational fields. Such a generalization
faces two immediate issues. First, by virtue of the existence of asymptotia,
the Cauchy slices must be {\em noncompact}. 
Second, by virtue of asymptotic flatness, the spatial triad must asymptote
to a {\em smooth} flat triad which is at odds with the discrete, non-smooth
spatial geometry alluded to above.

We shall return to a discussion of the first issue in the concluding section of this
paper. Here,  let us focus on the second issue ignoring complications arising from non- compactness.
It is expected that the effective smoothness
of classical geometry arises through coarse graining of its LQG quantum 
counterpart. Thus one would expect that the asymptotic conditions
translate to a requirement on suitably defined coarse grained properties
of the quantum states. While a final understanding of the quantum states
underlying asymptotically flat geometries would require such a treatment,
as a first step, it is useful to enquire, already in the case of compact spatial topology,
 if there is some way in which the 
standard LQG representation can be modified so as to directly accomodate
smooth spatial triads at the quantum level without explicit coarse graining.
An affirmative answer to this query is provided by the representation 
constructed by Koslowski in his seminal contribution \cite{tk}.
% at the kinematic level
% and further studied by  Sahlmann \cite{hs} at the $SU(2)$ gauge invariant
%and diffeomorphsim invariant level wherein group averagi.....
%The Koslowski- Sahlmann (KS) 
This representation 
assigns an extra label ${\bar E}^a_i$,
%where 
to the standard kinematic LQG states.
Here ${\bar E}^a_i$ is a smooth triad field. 
Triad dependent operators acquire 
an extra (smooth) contribution from ${\bar E}^a_i$ in addition to the 
standard discrete contributions \cite{tk,hs}. While Koslowski  restricted 
attention to
nondegenerate ${\bar E}^a_i$, this representation admits a straightforward
and obvious generalization to triads of arbitrary (and, in general,  spatially varying) degeneracy as well, and, indeed,
the ${\bar E}^a_i= 0$ sector of this representation turns out be exactly the
standard LQG one.

In a putative generalization of the representation 
to the asymptotically flat case it  would perhaps be appropriate to
retain standard  LQG structures in the interior while 
capturing flatness at asymptotia.
This would necessitate the consideration of states which are labelled 
by electric fields which are flat at asymptotia and vanish
in the interior (more precisely, inside of compact sets).
Hence, in anticipation of such a generalization of the representation to the asymptotically flat case, 
it is of relevance to study
its properties in the compact case  in the context of ${\bar E}^a_i$ of varying degrees
of degeneracy. 

As mentioned above such a study is trivial at the level of the kinematic
representation. However, physically relevant configurations are those
which are invariant under the action of the $SU(2)$ gauge group, as well
as the action of spatial diffeomorphisms and the Hamiltonian constraint.
Since the construction of the latter is an open issue even in standard LQG
we are interested, as a first step, 
 in imposing the diffeomorphism and $SU(2)$ gauge constraints
in this representation in the context of `background' electric fields
${\bar E}^a_i$ of spatially varying degeneracy.
Sahlmann initiated the imposition of gauge and diffeomorphism invariance 
via an application of group averaging techniques to this representation
in his pioneering work of Reference
\cite{hs} wherein he restricted attention to the case of non-degenerate triads. 
Accordingly, the first aim of this work is to initiate an investigation into the group averaging of
triads of spatially varying degeneracy.
%The purpose of this work is to build on Sahlmann's work
%using a key insight into the structure of this
%representation discovered
%recently \cite{mv13}. 
%While we work here in the contex of spatially compact 
%topology, we view this as a necessary prerequisite to the 
%a generalization (now in progress) to the asymptotically flat case of 
%noncompact spatial topology.

In Reference \cite{mv13}, it was noted that this representation studied by Koslowski and Sahlmann (henceforth referred to as the Koslowski- Sahlmann or KS 
representation) supported,
in addition to the action of the standard holonomy- electric flux  operators,
operator correspondents of certain connection dependent functions, called
`background exponentials'. 
In this work we show that in  order to implement the gauge transformation 
properties of these new functions in quantum theory, 
the action of quantum $SU(2)$ gauge transformations of Reference \cite{hs}
needs to be augmented with a phase factor. 
While this augmentation can be
absorbed in a redefinition of states labelled by the nondegenerate triads
of Sahlmann's work\footnote{There may  exist exceptional geometries for which this is not true,  see section \ref{sec4B}.},  in the degenerate case (of rank 1) this is no longer true. 
Since the $SU(2)$ Gauss Law and the spatial diffeomorphism constraints
generate a gauge group $\G \rtimes \diff$ which is the semidirect product  of the group of finite $SU(2)$ 
transformations $\G$ with the group of  spatial diffeomorphisms $\diff$, 
these phases need to be 
understood when both sets of constraints are imposed.
Accordingly, the second aim of this work is to initiate an investigation into the structure and role of these
phases in the group averaging procedure.

The necessity of extra phase factors can be easily seen in a $U(1)$ version
of the KS representation.\footnote{This example is worked out in detail in section \ref{sec4A}.}
 Consider any $U(1)$ gauge invariant state in the 
standard LQG type representation for gauge group $U(1)$ and augment it with 
a background electric field label ${\bar E}^a$. For concreteness
let the standard LQG type state be a $U(1)$ spin network. Then, 
$U(1)$ gauge transformations do not change the spin network
labels because these are anyway gauge invariant and do not change
the label ${\bar E}^a$ because electric fields are $U(1)$ gauge invariant.
If we ignore the phase factors alluded to above, the state is left invariant
by any $U(1)$ gauge transformation and hence should be annihilated by the 
Gauss Law constraint. But this is not true because the Gauss Law constraint
being just the divergence of the electric field operator, yields the 
divergence of ${\bar E}^a$ when it acts on the state in question, and $\Eb^a$ may be chosen so that its divergence is non-vanishing! When the phase factors are included
it turns out that the action of finite gauge transformations rephase the 
state with a phase proportional to the divergence of ${\bar E}^a$. The group 
averaging procedure averages over these phases and produces a vanishing 
result due to phase factor cancellation unless ${\bar E}^a$ is divergence free
in which case the state is exactly invariant under gauge transformations 
as expected.
Thus, to ensure correct results, it is crucial to keep account of the delicate
phases alluded to above. 

Let us return to the case of interest, namely that of gauge group
$\G \rtimes \diff$ with  $\G$ being the group of internal $SU(2)$ transformations.
The KS Hilbert space is spanned by an orthonormal basis of `KS' spin net states which are
generalizations of the standard LQG spin net states.  Each KS spin net is specified  by  
an $SU(2)$ gauge invariant spin net label `$s$' together with a background field label $\Eb^a_i$.\footnote{In section \ref{sec9B} we comment on the possibility of allowing for gauge variant spin networks.}
%In particular, there exist `pure background' KS states which are 
%generalizations of the cyclic `trivial' graph state of standard LQG.
%In order to 
%study novel features of the group averaging procedure which arise due to the key feature of the KS representation
%which is not shared by standard LQG, namely the existence of the background  triad label, it is useful to
%first study the 
%group averaging of these `pure background' states with backgrounds of spatially varying rank.
The group averaging procedure 
applied to a KS spin net state seeks to construct a corresponding $\G \rtimes \diff$ invariant state
as an `average' over $\G \rtimes \diff$ related images of the spin net. Recall that in the standard LQG
case,
an important role is played by transformations  which leave this spin net 
invariant. More in detail, given an $SU(2)$ invariant spin network $s$ these `symmetries' -corresponding to  diffeomorphisms leaving the spin net label $s$ invariant- 
determine the {\em superselection} sector containing the spin net state $|s\rangle$
\footnote{States in a single supereselection sector are mapped to each other by an appropriately defined
set of gauge invariant observables whereas states in distinct superselection sectors cannot be mapped
to each other by any such observable.
}
as well the detailed group averaging map in each  superselection sector. In particular (see \cite{almmt,Tbook}
as well as section IV of this paper), the structure of these symmetries implies that LQG spinnets living on 
graphs which are not related by diffeomorphisms lie in distinct superselected sectors and that 
within each such sector the group averaging map is well defined and unambiguous. Therefore, we expect that 
in the KS case too the symmetries of the label set of the KS spin net being averaged play similar, key role
in the averaging procedure. Accordingly, a large part of our analysis is focussed on trying to understand
the symmetries of these labels. Since one of these labels is a {\em field} with 3 dimensional support (in contrast
to the LQG case wherein the graph label is a 1 dimensional set), these symmetries are (infinitely!) more 
complex than those encountered in LQG. As a result, explicit results of a general nature are hard to come by.

After these remarks, let us now summarize our main results.
Our first set of results pertain to superselection criteria. Similar to the support of the spinnet 
graph being associated with superselection in LQG, here we show that appropriately defined support sets 
of the background field serve as (partial) superselection labels. Specifically, let the background field 
have rank 0, 1 or (2,3) in the `support' sets $V_0, V_1$ and $V_2$. Then, if two KS states have support sets (upto
sets of zero measure) which are unrelated by the action of any diffeomorphism, their group averages 
lie in distinct superselection sectors. 
%As seems plausible, a key role is played by the support sets of constant rank of the background field
%under consideration. 
We note here that {\em semianalyticity} of the background field plays a key role in our analysis. More in detail,
these support sets  may be written in terms of the zero sets of appropriate 
functions of the background field. Now, it is well known that zero sets of smooth functions 
do not, in general, have any nice properties. Hence it would be very hard, if not impossible, to
proceed with our analysis for {\em smooth} (i.e. $C^{\infty}$) background fields. However, standard LQG is 
most elegantly formulated for the {\em semianalytic} category \cite{lost,fleischhack}. 
In particular, the spatial diffeomorphisms considered in standard LQG are semianalytic and 
preserve semianalyticity as opposed to smoothness. Hence it is necessary for the KS background fields
to also be semianalytic. It turns out that the zero sets of semianalytic functions on compact manifolds have very nice 
properties, namely they are constructed as  finite unions of semianalytic
submanifolds \cite{lost,milman}. It is this beautiful property which allows us to 
perform a fairly detailed analysis and derive the above superselection result.

Our second set of results pertain to the role of phase contributions in the averaging procedure.
As in standard LQG, due to the absence of a well defined group invariant measure, our starting point
is the definition of a putative group averaging map as a formal sum over gauge related states.
Our results are as follows. First we show that if a KS state can be non-trivially rephased by any gauge transformation,
its image under the putative averaging map vanishes. Next, we show that  such phasings may manifest
for non-degenerate triads with appropriate symmetries and that such phases {\em do}
manifest {\em generically} if the rank 1 support set is of non- zero measure.

Our third set of results pertain to the case where the rank 1 support set is of zero measure
so that  the triad is exclusively of rank 0 or (2,3) almost everywhere. We show that, modulo one 
reasonable assumption (which needs to be proved), the group averaging map in this superselection 
sector is well defined and unambiguous.

Besides these fairly general results, we also derive a variety of less general, `case by case' results
when the rank 1 set is of non- zero measure. Our hope is that the material in this paper can serve
as a starting point for further studies of the complex and structurally rich problem of group averaging for
this (i.e. rank 1 support set of non- zero measure) case.

The layout of this paper is as follows. Section \ref{sec2} is devoted to a review of the necessary
background for our considerations. This includes:(a) the definition of classical holonomy, flux and 
background exponential functions, (b) their transformations under  
the gauge transformations generated by the 
classical $SU(2)$ Gauss Law and spatial diffeomorphism constraints of gravity, (c) a demonstration \cite{mv13}
that the KS Hilbert space provides a representation for the operator correspondents of the functions in (a)
as well a unitary representation of the transformations of (b). In particular, we highlight the 
additional phase contributions alluded to above.
%Section 2B and 2C review the material in References \cite{tk,hs,mv13} and Section 2D reviews the 
%group averaging procedure as outlined in, for example Reference \cite{almmt}.
In section \ref{sec3} we outline our general strategy for the construction of the group averaging map, emphasizing the 
importance of appropriate phase contributions in its construction so as to anticipate the corresponding 
subtleties to be encountered in the following sections. In particular we show that states which admit 
non-trivial rephasings average to zero. In \ref{sec4} we derive the phasing related results alluded to above
for nondegenerate triads with  symmetries and for generic triads with rank 1 sets of non- zero measure.

In section \ref{sec5}, we re- derive the standard LQG group averaging map from a slightly different 
perspective to the standard one. Our treatment emphasizes  the relative roles of various symmetry structures 
in the group averaging process and serves as a preview for the considerations which inform sections
\ref{sec6}- \ref{sec8}. In section \ref{sec6} we derive our principal result on superselection 
in the KS representation, namely its dependence on the support sets of rank 0, 1 and (2,3) of the background triad
label. In section \ref{sec7} we show 
(modulo certain reasonable assumptions) the existence of a well defined and ambiguity free
group averaging map for superselection sectors labelled by rank 1 sets of zero measure.
Section \ref{sec8} is devoted to a discussion of the various subtelities associated with the case when 
the rank 1 set is not of zero measure.
Section \ref{sec9} is devoted to a discussion of our results, open questions and to an account of work in progress.
Assorted technical details are collected in the appendices.

In this work we use units such that $c= 8\pi \gamma G =\hbar =1$ where $\gamma$ is the Barbero- Immirzi
parameter. Further, all differential geometric structures of interest will be based on the semianalytic, $C^k,k \gg 1$
category (see Appendix \ref{saapp} for further details). Finally, the Cauchy slice $\Sigma$ is assumed to be 
a compact without boundary semianalytic manifold.

\section{Preliminaries} \label{sec2}
The standard LQG quantization  (prior to imposition of
$SU(2)$ gauge and spatial diffeomorphism invariance) provides a representation of the algebra generated by 
holonomies (more precisely, their matrix components)  and electric fluxes. As shown in Reference \cite{mv13}, 
the KS quantization generalises
this standard LQG one so as to provide a representation of the enlargement of the holonomy- flux algebra to 
one which includes certain connection dependent functions referred to as `background exponential' functions.
As discussed in section \ref{sec1}, the gauge symmetry group of interest is the semidirect product of the 
group of local $SU(2)$ rotations with that of spatial diffeomorphisms.
Accordingly, 
in section \ref{sec2A} we review the classical phase space underlying LQG and  the definition of the holonomy, flux
and background exponential functions thereon. In section \ref{sec2B} we describe   the action of the gauge symmetry group
on the phase space and its induced action on the newly introduced background exponential functions.
In section \ref{sec2C} we review the KS representation of the enlarged algebra of holonomies, fluxes and background
exponentials and display a unitary representation of the gauge symmetry group in this representation.

We shall use the following nomenclature. Structures used prior to the imposition of the gauge
symmetry group are referred to as {\em kinematic}. Local $SU(2)$ gauge transformations are referred to as
{\em internal gauge} transformations. The gauge symmetry group will be referred to as the {\em bundle automorphism}
group \cite{hs} or, in short, as the {\em automorphism} group.

\subsection{Classical phase space and functions thereon} \label{sec2A}

The  classical phase space is coordinatized by  an $SU(2)$ connection $A_a$, and its conjugate
(unit density weight) $su(2)$ valued  electric field $E^a$ on the Cauchy slice $\Sigma$. In terms of their components in an $su(2)$ basis $\t_i, i=1,2,3$ with $[\t_i,\t_j]=\epsilon_{i j k} \t_k$,  $A_a=A_a^i \t_i$ and $E^a= E^a_i \t_i$, their Poisson brackets read: $\{A_a^i(x), E^b_j(y)\} = \delta^i_j \delta^b_a \delta (x,y)$.   We define the following functions on phase space:
\begin{eqnarray}
h_{e }(A) &:= & \P e^{\int_{e} A} , \label{hol}\\
F_{S,f}(E) &:=& \int_S dS_a \tr[f E^a], \label{flux}\\
\beta_{\Eb}(A) &:= &e^{i \int_\Sigma \tr[\Eb^a A_a]}. \label{baexp}
\end{eqnarray}
Here $h_{e }(A)$ is the $SU(2)$ matrix valued holonomy of the connection along the one dimensional oriented curve  or `edge' 
$e \subset \Sigma$. The normalization of the trace $\tr$ is taken so that $\tr[\t_i \t_j]=\delta_{ij}$.  $F_{S,f}(E)$ is the electric flux smeared with the $su(2)$-valued function $f$ through the surface
$S\subset \Sigma$. $\beta_{\Eb}(A)$ is a function which only depends on the connection and is obtained
by exponentiating the integral of the connection smeared with an $su(2)$ valued unit density weight vector
field ${\bar E}^a$. We refer to the c-number smearing function ${\bar E}^a$ as a background electric field
and to  $\beta_{\Eb}(A)$ as a background exponential.%\footnote{The manifold $\Sigma$ will be assumed to be compact without boundary. As indicated at the end of section 1, $\Sigma$ and its diffeomorphisms, as well as the edges $e$, surfaces $S$, internal gauge transformations, and the smearing functions $f,{\bar E}^a$ of equations (\ref{flux}),(\ref{baexp}) will all be of the semianalytic, $C^k, k \gg 1$ category.}

The Poisson brackets involving the background exponentials are:\footnote{In the second equation of  (\ref{pbbeta}),  $F_{S,f}$ inside the 
brackets is a phase space function, whereas  $F_{S,f}(\Eb)$ is  just the number  
$F_{S,f}(\Eb) = \int_S dS_a \tr[f \Eb^a]$.}
\be
\{ \beta_{\Eb},\beta_{\Eb'} \}=\{ \beta_{\Eb}, h^{}_{e} \} =0, \quad  \{ \beta_{\Eb}, F_{S,f} \} = i F_{S,f}(\Eb)
\beta_{\Eb}. \label{pbbeta}
\ee

\subsection{The group of gauge symmetries}\label{ggs} \label{sec2B}

Let the group of internal $SU(2)$ gauge tranformations be $\G$ and that of spatial diffeomorphisms be $\diff$,
both groups consisting of transformations connected to identity. Denote the gauge symmetry group, referred to by
Sahlmann as the bundle automorphism group  by $\aut$ so that  $\aut:=\G \rtimes \diff$. Elements of $\aut$ will be denoted as
 $a= (g,\phi) \in \aut$ with $g\in \G$ and $\phi \in \diff$. The product structure is given by (see appendix \ref{appsd} for further details):
\be
(g, \phi)  (g', \phi') =( g \phi_* g', \phi \circ \phi') \label{aap},
\ee
where $\phi_*$ denotes the push-forward action so that $(\phi_* g)(x)=g(\phi^{-1}(x))$. Infinitesimal generators of $\aut$ will be denoted by $(\Lambda,\xi)$ with $\Lambda$ an $su(2)$ valued scalar and $\xi$ a vector field. The infinitesimal version of (\ref{aap}) corresponds to the  commutator:\footnote{In the present setting of $C^k$ fields, the infinitesimal generators do not form a Lie algebra since the Lie bracket of two $C^k$ vector fields is in general $C^{k-1}$.  But they do generate well defined  one parameter subgroups   of $\aut$. }
\be
[(\Lambda,\xi),(\Lambda',\xi')]  = (-\L_\xi \Lambda' + \L_{\xi'} \Lambda+[\Lambda,\Lambda'], [\xi,\xi']),
\ee
where $[\Lambda,\Lambda']$ is the $2 \times 2$ matrix commutator and $[\xi,\xi']$ the vector field Lie bracket. 
The group $\aut$ acts on phase space according to:
\be
\begin{array}{lll}
(g, \phi) \cdot A_a & :=& g \phi_* A_a g^{-1} - (\partial_a g) g^{-1}  ,\\
(g, \phi) \cdot E^a & := & g  \phi_* E^a g^{-1}, \label{autAE}
\end{array}
\ee
with the corresponding  infinitesimal version:
\be
\begin{array}{lllll}
(\Lambda,\xi) \cdot A_a & :=& [\Lambda,A_a] - \L_\xi A_a -\partial_a \Lambda &=& \{  G[\Lambda]+D[\xi], A_a \}, \\
(\Lambda,\xi) \cdot E^a &:=& [\Lambda,E^a]- \L_\xi E^a &=& \{  G[\Lambda]+D[\xi], E^a \} , \label{lieautAE}
\end{array}
\ee
generated by Poisson brackets with the  Guass law and diffeomorphism constraints
%\footnote{Our conventions are such that, for any 2 by 2 matrix $\Lambda$, $\tr[\Lambda]:=-2 \text{Trace}[\Lambda]$  
%so that the  $\tr[\tau_i \tau_j]=\delta_{ij}$ for the standard $su(2)$ antihermitian generators satisfying  
%$\tau_i \tau_j=\tfrac{1}{2}\e_{ijk}\tau_k -\tfrac{1}{4}\delta_{ij} \idtwo$. We work in units where 
%$8 \pi G \gamma= c= \hbar =1$. The fundamental Poisson brackets are: 
%$\{ A_a^i(x),E^b_j(y) \}=\delta_a^b \delta^i_{j} \delta(x,y)$.}
%
\ba
G[\Lambda]& = &-\int_\Sigma \tr[\Lambda (\partial_a E^a+[A_a,E^a] )] ,\label{gausslaw}\\
D[\xi]&=& \int_{\Sigma}  \tr[E^a \L_{\xi}A_a ].
\ea
The action of the group $\aut$ on the phase space functions (\ref{hol}), (\ref{flux}), (\ref{baexp}) is given by:
\ba
a \cdot h_e(A) &:=&  h_e(a^{-1} \cdot A) \nn \\
&=&g^{-1}(\phi(e_i))h_{\phi(e)}(A)g(\phi(e_f)) , \label{ahol}\\
a \cdot F_{S,f}(E) &:=& F_{S,f}(a^{-1} \cdot E) \nn \\
&=& F_{\phi(S),g \phi_* f g^{-1}}(E) , \label{aflux}
\ea
where $e_i$ and $e_f$ denote the initial and final points of the edge $e$. The transformation law for the  background exponentials is found to be
\ba
a \cdot \, \bE [A] & :=& \bE[a^{-1}\cdot A] \nn \\
& =& e^{i \alpha(a,\Eb) }\beta_{a \cdot \Eb}[A], \label{abaexp}
\ea
where
\be
\alpha(a,\Eb):= \int_\Sigma \tr[\phi_* (\Eb^a) g^{-1}\partial_a g]. \label{alpha}
\ee
By construction, (\ref{ahol}), (\ref{aflux}) and (\ref{abaexp}) provide a representation of $\aut$ on holonomies, fluxes and background exponentials. Since this fact may  not be immediately obvious for the newly introduced functions,  let us explicitly verify that  (\ref{abaexp}) satisfies
\be
a\cdot (a' \cdot  \bE)=(a  a')\cdot  \bE .\label{a1a2beta}
\ee
The left and right hand side of (\ref{a1a2beta}) are given by:
\ba
a\cdot (a' \cdot  \bE) &=& e^{i\alpha(a,a' \cdot \Eb)}e^{i \alpha(a',\Eb)}\beta_{a \cdot (a' \Eb)} \label{lhsaa}\\
(a   a')\cdot  \bE&=& e^{i \alpha(a a',\Eb)}\beta_{(a a') \cdot \Eb} \label{rhsaa}.
\ea
Since $a \cdot (a' \Eb)=(a a') \cdot \Eb$, all we need to check is that the phases in (\ref{lhsaa}) and (\ref{rhsaa}) agree. Starting from the phase in (\ref{rhsaa}) for $a=(g,\phi), a'=(g',\phi')$ and using  (\ref{aap}) and (\ref{alpha}), we find:
\ba
\alpha(a a',\Eb)& =& \int_\Sigma \tr[(\phi \circ \phi')_* (\Eb^a) (\phi_* g'^{-1} g^{-1})\partial_a (g \phi_* g')] \nn \\
&=&\int_\Sigma \tr[\phi_*(  \phi'_* \Eb^a g'^{-1}) g^{-1}\partial_a g \phi_* g'] +
\int_\Sigma \tr[\phi_*(  \phi'_* \Eb^a) \phi_* g'^{-1}  \phi_* \partial_a g'] \nn \\
&=&\int_\Sigma \tr[\phi_*( g' \phi'_* \Eb^a g'^{-1}) g^{-1}\partial_a g ] +
\int_\Sigma \tr[ \phi'_* \Eb^a  g'^{-1}   \partial_a g'] \nn \\
&=&\alpha(a,a'\cdot\Eb)+\alpha(a',\Eb). \label{phasesaap}
\ea
 %where we used cyclicity of the trace, invariance of the integral under  diffeomorphisms, and $\partial_a \phi_*g'=\phi_* \partial_a g'$. 
 Thus, the phases  in (\ref{lhsaa}) and (\ref{rhsaa}) agree  and  (\ref{a1a2beta}) is indeed satisfied.

We conclude the section by describing a key  property of the phase factors. Given a background electric field $\Eb$, we define its symmetry group by:
\be
\S_{\Eb }:= \{ a \in \aut \; : \;  a \cdot \Eb  =   \Eb \} \subset \aut. \label{symE}
\ee
Consider the map $\pi_{\Eb} : \S_{\Eb} \to U(1)$ defined by
\be
\pi_{\Eb}(a):= e^{i \alpha(a,\Eb)} , \quad a \in \S_{\Eb}.\label{defpi}
\ee
Thus, $a \cdot \beta_{\Eb}= \pi_{\Eb}(a) \beta_{\Eb}$ for $a \in \S_{\Eb}$. From the property (\ref{a1a2beta}) it follows that $\pi_{\Eb}(a_1) \pi_{\Eb}(a_2)=\pi_{\Eb}(a_1 a_2)$ for $a_1,a_2 \in \S_{\Eb}$, i.e. $\pi_{\Eb}$ is a group homomorphism form  $\S_{\Eb}$ into $U(1)$.

\subsection{The KS representation} \label{KSkin} \label{sec2C}
The kinematical Hilbert space of standard LQG is spanned by the orthormal basis of spin network states $\{|s\rangle\}$.
Let the dense domain of the finite linear span of spinnets be ${\cal D}$. Let $\hat O$ be  an operator 
from ${\cal D}$ to ${\cal D}$ so that  $\hat O|s\rangle$ is a finite linear combination of spinnets i.e.  
$\hat O|s\rangle = \sum_IO^{(s)}_I|s_I\rangle$ where $O^{(s)}_I$ are the complex  coefficients in the sum over the spinnets
$|s_I\rangle$. 
It is useful to introduce the notation $|{\hat O}s\rangle$ to denote this linear combination of spinnets so that 
we have 
\begin{equation} 
|{\hat O}s\rangle:= \hat O|s\rangle = \sum_IO^{(s)}_I|s_I\rangle .
\label{o}
\end{equation}

The KS Hilbert space is then spanned by states  which have, in addition to their LQG
spinnet label, an  additional label $\Eb^a$ where ${\Eb^a}$ is a background electric field.
We denote such a state by $|s,{\Eb}\rangle$. These 
states for all $s, \Eb^a$ provide an orthonormal basis for the Koslowski- Sahlmann kinematic Hilbert space
so that the inner product between two such KS spinnets in this Hilbert space is
\begin{equation}
\langle s^{\prime},{\Eb}^{\prime }_{}|s,{\Eb}\rangle = \bra s|s^{\prime} \ket
\delta_{{\Eb}^{\prime}, {\Eb}}, \label{ksip}
\end{equation}
where $\bra s|s^{\prime} \ket$ is just the standard LQG inner product and the second factor is the Kronecker delta which 
vanishes unless the two background fields agree in which case it equals unity.

The holonomy- flux operators act on the KS spinnets as:
\begin{eqnarray}
{\hat h}^{\phantom{e\;}A}_{e\;B}|s,\Eb \rangle &:=& \vert {\hat h}^{\lqg \,A}_{e\; \quad B} s,\Eb_{}\rangle ,
\label{holhat}
\\
{\hat F}_{S,f}|s,\Eb \rangle &:=& |{\hat F}^\lqg_{S,f} s,\Eb_{}\rangle + F_{S,f}(\Eb)|s,\Eb_{}\rangle .
\label{fluxhat}
\end{eqnarray}
Here
${\hat h}^{\;A}_{e \;B}$
is the $A,B$ component of the holonomy operator.  
Since the holonomy of equation (\ref{hol})
is in the defining $j=\frac{1}{2}$ representation, the indices  $A,B$ take values  in $\{1,2\}$ . 
Further, we have used the notation of equation (\ref{o}) so that 
 ${\hat h}^{\lqg \,A}_{e\; \quad B}s$ and ${\hat F}^\lqg_{S,f} s$  represent the standard LQG action on spin networks.
The background exponential operators act through:
\be
{\hat \beta}_{\Eb'} |s,\Eb \ket : =  |s, \Eb'+\Eb \ket \label{betahat}.
\ee
It then follows that (\ref{holhat}) and (\ref{fluxhat}) satisfy the standard holonomy-flux commutation relations. It is easy to verify that the only additional non-trivial commutator is given by
\be
 [ {\hat \beta}_{\Eb}, {\hat F}_{S,f} ] = - F_{S,f}(\Eb) {\hat \beta}_{\Eb}, \label{commbeta}
\ee
in agreement with the  Poisson bracket in (\ref{pbbeta}).

The unitary action of the gauge group $\aut$ on the  KS Hilbert space is dictated by the transformation properties of the elementary phase space functions and reads:
\be
U(a) |s, \Eb \ket := e^{i \alpha(a,\Eb)} |U^{\lqg}(a) s, a \cdot \Eb \ket ,\label{UasE}
\ee
where $U^{\lqg}(a) s$ denotes the usual action of $\aut$ on spin networks, and $\alpha(a,\Eb)$ and $a\cdot\Eb$  are given in (\ref{alpha}) and (\ref{autAE})  for $a=(g,\phi)$.  It is immediate to verify that (\ref{UasE}) preserves the inner product (\ref{ksip}).
Further,  
from the fact that $U^{\lqg}(a_1 a_2) =U^{\lqg}(a_1)U^{\lqg}(a_2) $ in conjunction with  equation (\ref{a1a2beta}) it
follows  that $U(a_1 a_2)=U(a_1) U(a_2)$ so that equation (\ref{UasE}) defines a unitary representation of $\aut$.

Finally, we verify that (\ref{UasE}) reproduces the transformation rules (\ref{ahol}), (\ref{aflux}) and (\ref{abaexp}). 
For the holonomy operators (\ref{holhat}), this follows from the transformation rule being satisfied in the standard LQG space.
For the flux operators (\ref{fluxhat}) we have for $a=(g,\phi)$:
\ba
U(a) {\hat F}_{S,f} U(a^{-1}) |s,\Eb \ket &=& e^{i \alpha(a^{-1},\Eb)} U(a) {\hat F}_{S,f}  |U^{\lqg}(a^{-1}) s, a^{-1} \cdot \Eb \ket  \nn \\
&=&  e^{i \alpha(a^{-1},\Eb)} U(a) (|{\hat F}^\lqg_{S,f} U^{\lqg}(a^{-1}) s, a^{-1} \cdot \Eb \rangle + F_{S,f}(a^{-1} \cdot \Eb)|U^{\lqg}(a^{-1}) s,a^{-1} \cdot \Eb\rangle) \nn \\
&=&|{\hat F}^\lqg_{\phi(S),g \phi_* f g^{-1}}  s,  \Eb \rangle+ F_{\phi(S),g \phi_* f g^{-1}}(\Eb) | s,  \Eb \rangle \nn\\
&=&{\hat F}_{\phi(S),g \phi_* f g^{-1}}|s,\Eb \ket, \label{uaflux}
\ea
where in going from the second to third line we used the transformation rule for the standard LQG flux, as well as the following cancellation of phases:
\be
e^{i \alpha(a^{-1},\Eb)}e^{i \alpha(a, a^{-1}\cdot \Eb)} =1 ,\label{phasecancel}
\ee
as follows from (\ref{phasesaap}) by setting  $a'=a^{-1}$. For the  background exponential operators (\ref{baexp}) we have:
\ba
U(a) {\hat \beta}_{\Eb'} U(a^{-1}) |s,\Eb \ket & = & e^{i \alpha(a^{-1},\Eb)}U(a) {\hat \beta}_{\Eb'}  |U^{\lqg}(a^{-1}) s, a^{-1} \cdot \Eb \ket  \nn \\
&= & e^{i \alpha(a^{-1},\Eb)}U(a)  |U^{\lqg}(a^{-1}) s, \Eb'+a^{-1} \cdot \Eb \ket \nn \\
&= & e^{i \alpha(a^{-1},\Eb)}e^{i \alpha(a, \Eb'+a^{-1}\cdot \Eb)}  |s, a \cdot \Eb'+  \Eb \ket \nn \\
&=&e^{i \alpha(a, \Eb')} |s, a \cdot \Eb'+  \Eb \ket \nn \\
&=&e^{i \alpha(a, \Eb')} {\hat \beta}_{a \cdot \Eb'}|s,   \Eb \ket \label{uabaexp}
\ea
where in going from the third to fourth line we  used (\ref{phasecancel}). It follows from equations 
(\ref{uaflux}) and  (\ref{uabaexp})   
that the unitary representation  (\ref{UasE}) reproduces the transformation rules (\ref{aflux}), (\ref{abaexp}) for the flux and background exponential operators.

We conclude by pointing out that the actual kinematical space to be used in  the text is the subspace generated by states $|s,\Eb \ket$ such that  $s$ is $SU(2)$ gauge invariant. This is  the Hilbert space to be  referred  as `kinematical' and  denoted by  $\Haux$.  We comment on the reasons for this  restriction in section \ref{sec9B}.
\section{Outline of our general strategy} \label{sec3}

In section \ref{sec3A} we review the defining properties of  any satisfactory group averaging map.
Any such map yields gauge invariant states from appropriate kinematic ones.
Following References \cite{almmt,hs}, a candidate for such a map may be constructed as a formal
sum over all distinct `bras' which are  gauge related to the kinematic one under consideration,
the idea being that the action of any gauge transformation on this sum sends the sum into itself
thus ensuring gauge invariance. In section \ref{sec3B} we 
organize this (formal) sum in a way which  anticipates   
the context of the 
unitary representation of the 
group $\aut$ described in section \ref{sec2}, with particular emphasis on 
the structure and effect of phase contributions of the type encountered in equation (\ref{UasE}). The arguments we use to organise the sum are of a slightly formal nature and  
framed in a general context. Their purpose is to serve as heuristic motivation for the 
group averaging maps constructed in detail in the context of the KS representation. 
In section \ref{sec3C} we note that similar to the case of LQG \cite{almmt}, 
when applied to KS states, our strategy yields a putative averaging map with infinitely many
unknown positive definite parameters. As in
Reference \cite{almmt} we recall the relevance of the phenomenon of superselection
in evaluating the import of these parameters and 
show how the ambiguities in their values
can be  reduced  using properties of gauge invariant observables.
Finally, in section \ref{sec3D} we use the material presented in sections \ref{sec3A} to \ref{sec3C} to  
formulate the strategy followed in sections \ref{sec5} to \ref{sec7}.

\subsection{Defining properties of a Group Averaging map} \label{sec3A}

Let the gauge group $\aut$  of interest be represented unitarily on the kinematic Hilbert space
${\cal H}_{kin}$. A satisfactory Group Averaging map $\eta$ is an
anti linear map $\eta: \D \to \D'$,  from a dense domain $\D \subset {\cal H}_{kin}$  that is preserved under the unitary action of $\aut$  to the space
$\D'$ of complex linear mappings on $\D$ ($\D'$ is called the {\em algebraic dual} of $\D$), 
satisfying the following three properties \cite{almmt,raq}:
\begin{enumerate}
\item  $\forall \, \psi_1  \in \D$, $\eta(\psi_1) \in \D'$ is $\aut$-invariant:
\be
\eta(\psi_1)[U(a) \psi_2]= \eta(\psi_1)[\psi_2] \quad \forall \, a \in \aut  , \;\psi_2 \in \D \label{prop1}
\ee
\item $\eta$ is real and positive:
\be
\eta(\psi_1)[\psi_2]  = \overline{\eta(\psi_2)[\psi_1]}  \; ,  \quad \eta(\psi_1)[\psi_1] \geq  0 \;  \quad \forall \, \psi_1, \psi_2 \in \D \label{prop2}
\ee
\item $\eta$ commutes with the observables:
\be
\eta(\psi_1)[O \psi_2]=\eta(O^\dagger \psi_1)[\psi_2] \quad \forall \, \psi_1,\psi_2 \in \D, \; \forall \,  O \in \O . \label{prop3}
\ee
\end{enumerate}
In the last condition  ``observables''  stand for ``strong observables that preserve $\D$ '', that is:
\be
\O :=\{ O : \D \to \D \; |\;  U(a)O=O U(a)\quad \forall a \in \aut \}.
\label{3.4}
\ee
Once one succeeds in finding such a map $\eta$, the  $\aut$-invariant Hilbert space $\Haut$ is obtained as follows \cite{raq}:  Let $\Vaut \subset \D'$  be the span of dual vectors of the form $\eta(\psi)$. The sesquilinear form  $\bra \eta(\psi_1) , \eta(\psi_2) \ket_{\text{Aut}}:=\eta(\psi_2)[\psi_1]$ provides an inner product on $\Vaut/\sim$ where the quotient is over zero-norm states. Property 2 implies it is an inner product, and  $\Haut$ is defined as the completion of $\Vaut/\sim$ under this inner product. Property 3 ensures that strong observables are well defined on $\Haut$, and satisfy the correct  adjointness relations on  $\Haut$ if they do so on $\D$ \cite{Tbook}.

\subsection{Group Averaging as a sum over states} \label{sec3B}

We write the $\aut$ invariant image of the kinematic state $\vert \psi\rangle$ as the formal sum:
\be
\eta (\vert \psi\rangle  ) = \sum_{|\phi\ket \in {\rm Orb} (\psi )}  |\phi\rangle^{\dagger}, 
\label{orbit}
\ee
where  ${\rm Orb} (\psi )$ is the orbit of $|\psi\ket$ under $\aut$ i.e. the set of all distinct gauge related images
of $|\psi \ket$, and we have used the notation $|\phi\rangle^{\dagger}:= \langle \phi \vert$. 
Every such element may be written as $|\phi\ket = { U}(a) |\psi\ket $ for some $a \in \aut$. 
Let us, in this manner,
arbitrarily choose one such element $a$  for each element in ${\rm Orb} (\psi )$  and call the resulting set
$\aut_{{\rm Orb}(\psi )}$ so that 
\be
\eta (|\psi\ket ) = \sum_{a \in \aut_{ {\rm Orb}(\psi  )}} (U(a)| \psi \ket )^{\dagger} .
\label{autorbit}
\ee
Elements of $\aut_{{\rm Orb}(\psi )}$ can be characterised as follows.
Let $\sym_{\psi}\subset \aut$ be the set of symmetries of $|\psi \ket$ i.e. the set of automorphisms which leave $|\psi \ket$
invariant so that $U(a)|\psi \ket =U(b)|\psi \ket$ iff $a= bs$ for some $s\in \sym_{\psi}$. This implies that elements of 
$\aut_{{\rm Orb}(\psi )}$ are in correspondence with the cosets of $\aut$ by $\sym_{\psi}$. 
Next, recall that 
elements of this coset space, $\aut /\sym_{\psi}$,  are the equivalence classes $[a]$ where $b \in [a]$
iff there exits some $s\in \sym_{\psi}$ such that $a= bs$. 
Using the defining properties of  $\aut_{{\rm Orb}(\psi )}$, the coset space $\aut /\sym_{\psi}$
and the symmetry group $\sym_{\psi}$, it
is easily verified that\\ (i) for any $b \in \aut$ and any $a_1,a_2 \in \aut_{{\rm Orb}(\psi )} \; a_1\neq a_2$,
it follows that $[ba_1]\neq [ba_2]$ \\
(ii)given any $c \in \aut_{{\rm Orb}(\psi )}$ and any $b\in \aut$, there exists
a unique $a \in \aut_{{\rm Orb}(\psi )}$ such that $[ba]= [c]$.\\
 It then follows from (i) and (ii) 
that for any $b\in \aut$:
\be
{U}^{\dagger}(b)\eta (|\psi \ket ) = 
\sum_{a \in \aut_{ {\rm Orb}(\psi )}} ({U}(b)U(a)|\psi \ket )^{\dagger}= 
\sum_{a \in \aut_{ {\rm Orb}(\psi )}} (U(ba) |\psi \ket )^{\dagger}
=\sum_{c \in \aut_{ {\rm Orb}(\psi )}} (U(c)|\psi \ket )^{\dagger},
\ee
thus establishing the formal gauge invariance of the sum.

Note that the above argument requires the sum to range over {\em all} distinct images of $|\psi \ket$. In particular,
images of $|\psi \ket$ which are distinct from each other but {\em proportional} to each other must be included in the
sum. Thus, if $|\phi \ket$ is in the sum and there exists $a\in \aut$ such that ${U}(a)|\phi \ket= c|\phi\ket, c\neq 1$, then 
$c|\phi\ket$ must also be in the sum. Note that $|c|=1$ by unitarity of  ${U}(a)$ so that $c = e^{i\theta}, \theta\in \reals$ 
is a phase factor. It is convenient for future purposes to define  the sum in equation (\ref{autorbit}) so as to sum
over such phase related states first. This is done as follows.

Let $\Ph_{\psi}\subset \aut$ be the set of gauge transformations which rephase $|\psi \ket$ so that ${U}(a)|\psi \ket$ and 
${U}(b)|\psi \ket$
are proportional iff $a=bp$ for some $p \in \Ph_{\psi}$. Thus, the set of `phase unrelated states' are
in correspondence with the coset space $\aut/ \Ph_{\psi}$ each coset consisting of elements of an 
equivalence class $[a]^{\Ph}$ where $a\equiv b$ iff $a=bp$ for some $p \in \Ph_{\psi}$. 
Let us choose, arbitrarily, one element from each such equivalence class and call the resulting set 
$\aut^{\perp}_{\Ph_{\psi}}$. It follows that every element of the orbit of $|\psi \ket$ can be obtained
by an appropriate rephasing of ${U}(a) |\psi \ket$ for some $a \in \aut^{\perp}_{\Ph_{\psi}}$.
Next, note that distinct rephasings of $|\psi \ket$ are in correspondence with cosets of 
$\Ph_{\psi}$ by the symmetry group $\sym_{\psi}$. It is easily verified that $\sym_{\psi}$ is a normal subgroup of
$\Ph_{\psi}$ so that the coset space $\Ph_{\psi}/\sym_{\psi}$ is just the quotient group obtained by 
quotienting $\Ph_{\psi}$ by $\sym_{\psi}$. It is then straightforward to see that 
$\Ph_{\psi}/\sym_{\psi}$ must be  homomorphic to $U(1)$ or a subgroup of $U(1)$.
In terms of this homomorphism we write the sum as:
\be
\eta (|\psi \ket ) = \left(\sum_{a \in \aut^{\perp}_{\Ph_{\psi}  }   }   
U(a) \sum_{e^{i\theta}\in  \Ph_{\psi}/\sym_{\psi}} e^{i\theta} |\psi \ket 
\right)^{\dagger} .
\label{phasesum}
\ee
If $\Ph_{\psi}/\sym_{\psi}$ is a non-trivial proper subgroup of $U(1)$ standard group theoretic results imply that
this subgroup is finite and that the sum over phases vanishes i.e. 
$\sum_{e^{i\theta}\in  \Ph_{\psi}/\sym_{\psi}} e^{i\theta}=0$. If $\Ph_{\psi}/\sym_{\psi}= U(1)$, we have an infinite
sum of phases which we can plausibly {\em define} to vanish by virtue of the fact that for every element
in the sum there is also its negative. Thus, in both these cases we have that $\eta (|\psi \ket )=0$.
Thus, the  only case for which the group averaging map could be non-trivial is when $\Ph_{\psi} =\sym_{\psi}$. The group averaging sum (\ref{autorbit}) then takes the form:
\be
\eta(|\psi \ket)= \left\{
\begin{array}{lll}
\quad \quad 0 & \text{if }  & \sym_{\psi} \subsetneq \Ph_{\psi} \\
& &\\
\displaystyle\sum_{a \in \aut_{ {\rm Orb}(\psi )}} (U(a) |\psi \ket )^{\dagger}  &\text{if } &\sym_{\psi} = \Ph_{\psi} 
\end{array} \right. 
\label{nophasesum}
\ee

In closing, we note that the first equation of (\ref{nophasesum}) must be satisfied by any well defined 
group averaging map. To see this, consider the case  where there exists $a \in \Ph_{\psi}$  that 
rephases $|\psi \ket$ by $e^{i\theta}\neq 1$. Then from the  gauge invariance condition (\ref{prop1}) we have that
\be
\eta(|\psi \ket)=\eta(U(a) |\psi \ket)= \eta(e^{i \theta} |\psi \ket)= e^{-i \theta} \eta(|\psi \ket), \label{phprop1}
\ee
which implies the vanishing of $\eta (|\psi \ket )$. From this point of view, the considerations in the main part of this 
section correspond to a particular mechanism for the vanishing of such $\eta (|\psi \ket )$.

\subsection{Ambiguities in the Group Averaging map}\label{sec3C}

Let $|\psi \ket$ be a KS spinnet and choose ${\cal D}$ (see section \ref{sec3A}) to be the finite linear span of KS spinnets. 
From equation (\ref{UasE}) it follows that 
every  state which is gauge related to $|\psi \ket$ is also a spinnet. To reduce notational clutter we shall 
call this set, referred to as ${\rm Orb} (\psi )$ in section \ref{sec3B}, as $[\psi]$.
It follows that if 
$\sum_{a \in \aut^{\perp}_{\Ph_{\psi}  }   }   
(U(a) |\psi \ket 
)^{\dagger}$ is a gauge invariant state, so is 
$\eta_{[\psi ]} \sum_{a \in \aut^{\perp}_{\Ph_{\psi}  }   }   
(U(a) |\psi \ket
)^{\dagger}$ for any constant  $\eta_{[\psi ]}$  (from property 2 (\ref{prop2}) this constant must be real and positive).

Recall that the dense set of gauge invariant states  ${\cal D}^{\prime}$ is obtained as the image of the set ${\cal D}$ 
(see section \ref{sec3A}). It then follows that every choice of the set of  coefficients $\eta_{[\psi ]}$, one for 
each gauge equivalence class of KS spinnets $[\psi]$, yields a putative group averaging map.
While some of the ambiguity in these choices can be absorbed into a rescaling of the gauge invariant Hilbert space
inner product (see section \ref{sec3A} 2.), a vast amibiguity still remains. As in the case of LQG \cite{almmt} we adopt the
view that such ambiguities are only of physical relevance within a single superselection sector of the 
gauge invariant Hilbert space. Recall that, roughly speaking,  if no observable  maps a set of states to its complement, 
we say that the set of states is superselected. In more detail,
using  the notation of section \ref{sec3A}, the notion of superselection sector is as follows.

Given two states $|\psi_1 \ket, |\psi_2 \ket \in \Haux$ we say  their corresponding $\aut$-invariant states $\eta(|\psi_i\ket)\in \Haut$ are superselected if
\be
\bra \eta( |\psi_1\ket) ,O \eta(|\psi_2\ket) \ket_{\text{Aut}}=0 \ , \forall \; O \in \O,   \label{ss}
\ee
or equivalently if, 
\be
\bra \psi_1 | O U(a) |\psi_2 \ket_{kin}=0 \quad \forall \; O \in \O, \; \forall a \in \aut.
\label{3.12}
\ee
The Hilbert space $\Haut$ decomposes then into superselected subspaces, each of which is left invariant under the action of all observables $O \in \O$. The viewpoint of Reference \cite{almmt} is that each superselection sector is
one possible realization of nature and it suffices to focus on one superselection sector at a time.
In particular we need only address the ambiguities in the choice of group averaging map within
the context of a single superselected sector. In practice, this reduces the choice of complex coefficients
$[\psi]$ drastically because overall scaling of the group averaging map in each superselection sector can be
reabsorbed into the Hilbert space inner product for that superselection sector.

Note that within a single superselection sector, obervables can (and do) map different gauge orbits $[\psi]$ to each
other. It then follows that the adjointness requirement 3. of section \ref{sec3A} yields consistency conditions 
between the ambiguity coefficients $\eta_{[\psi]}$ i.e. $\eta_{[\psi_1]}$, $\eta_{[\psi_2]}$ must be chosen
so as to satisfy requirement 3. of section \ref{sec3A} whenever the action of an observable  on a state in 
$[\psi_1]$ results in a state with overlap with state(s) in $[\psi_2]$. In practice, these consistency 
requirements often serve to remove the ambiguities in $\eta_{[\psi]}$ in a given superselection sector
\cite{pft,almmt}.

Finally, we show that states admitting rephasings are superselected from those that do not. Let $|\psi \ket$ be a state admitting rephasing so that there exists $a \in \aut$ such that $U(a) |\psi \ket = e^{i\theta} |\psi \ket$ with $e^{i\theta} \neq 1$. From the discussion of the previous section $\eta(|\psi \ket)=0$. Now, since our group averaging map satisfies  property 1, we can use the same reasoning as in Eq. (\ref{phprop1}) to conclude that $\eta(O |\psi \ket)$ vanishes for any observable $O$:
\be
\eta(O |\psi \ket)=\eta(U(a) O |\psi \ket)=\eta(O U(a)  |\psi \ket)= e^{-i \theta}\eta(O |\psi \ket).
\label{rephaseo}
\ee
This implies that  property 3. of the group averaging trivializes to  $0=0$ whenever one of the states admits rephasings. In particular  the superselection condition (\ref{ss}) will follow whenever one of the states admits rephasings. 

\subsection{Our Strategy}\label{sec3D}
Based on sections \ref{sec3A}-\ref{sec3C}, our strategy for the construction of a group averaging map for the KS representation
is as follows. We choose ${\cal D}$ to be the finite span of KS spinnets. We write the group averaging map 
applied to any KS spinnet $|\psi \ket$ in the form of the sum (\ref{nophasesum}) augmented with an ambiguity coefficient 
$\eta_{[\psi ]}$ i.e. 
\be
\eta(|\psi \ket)= \eta_{[\psi]}\displaystyle\sum_{a \in \aut_{ {\rm Orb}(\psi )}} (U(a)|\psi \ket )^{\dagger}  \quad \text{for } \sym_{\psi} = \Ph_{\psi} .
\label{finalsum}
\ee
 We verify that this sum defines an element of ${\cal D}^{\prime}$. We then isolate a large superselection
sector and reduce the ambiguity in the group averaging map when restricted to
such a sector by imposing requirement 3. of section \ref{sec3A}.

\section{Importance of phases} \label{sec4}

In this section we illustrate the importance and non trivial consequences of the presence of the phase term  $e^{i\alpha(a,\Eb)}$ (\ref{UasE}) in the unitary action of $\aut$ on the KS space. 

  We first analyze $U(1)$ gauge theory  to highlight the crucial role of the non- trivial 
rephasings of section \ref{sec3B} in obtaining the correct gauge invariant state space. We then move to the $SU(2)$ theory, where we show a close similarity of rank 1 triads with  abelian electric fields. We finally point out how phases may also have nontrivial effects in the case of  rank 2 or 3 triads.

\subsection{$U(1)$ Abelian example} \label{abeliancase} \label{sec4A}

Consider $U(1)$ gauge theory in a KS representation.  Since our purpose in this section is to provide the reader
with a setting wherein the importance of  phases  manifests in a direct and transparent manner,
we restrict attention below to a $\G$ (rather than  $\G \rtimes \diff$) averaging of  `pure background' sates of the form  $ |\Eb \ket:=|0,\Eb \ket$ , where $\Eb^a$  is now a single densitized vector field and ``$0$'' refers to the trivial graph spin network.

The gauge group $\G$ is now given by local $U(1)$ rotations $g=e^{i \theta}$ with $\theta : \Sigma \to \reals$. 
From equation (\ref{alpha}), the phase factor associated to $g$ evaluates to  
\be
\alpha(e^{i \theta},\Eb) = \int_\Sigma \Eb^a \partial_a \theta = -\int_\Sigma \theta \partial_a \Eb^a ,
\label{alphaabelian}
\ee
and the unitary action   (\ref{UasE}) becomes
\be
U(e^{i \theta}) |\Eb \ket = e^{-i \int_\Sigma \theta \partial_a \Eb^a } | \Eb \ket, \label{UaEabelian} 
\ee
where we have used the fact that $g\Eb g^{-1} = \Eb$.

Now, for a moment, let us see what happens if we ignore this phase factor so as to set 
$U(e^{i \theta}) |\Eb \ket = | \Eb \ket$ which  implies that $| \Eb \ket$ is gauge invariant. Thus if we did
this, all pure background states would be invariant under the group averaging procedure. This should be equivalent
to the statement that the Gauss Law is satisfied. As the reader may verify, in $U(1)$ theory, we are in 
the fortunate situation that 
the electric field operator is itself well defined and diagonalised by all KS spinnets (including, therefore,
pure background ones). One obtains $ \partial_a {\hat E}^a |\Eb \ket = \partial_a {\Eb}^a |\Eb \ket$ which vanishes as 
expected 
 only for divergence free electric fields.
This is in contrast to invariance imposed by the group averaging procedure without phasing which leads to the 
physically erroneous conclusion that all pure background states, whether 
labelled by divergence free electric fields or not, are gauge invariant!

Let us now use the correct quantum implementation of elements of $\G$ {\em with} 
the phasing of equation (\ref{UaEabelian}). From equation (\ref{alphaabelian}), the phase is nontrivial only
for $\Eb^a$ which is not divergence free. Then,
identical to the discussion just after equation (\ref{phasesum}), we have that the sum over such nontrivial phases
vanishes in the group averaging procedure. This implies that, as expected on physical grounds, 
the only nontrivial gauge invariant states are the  ones with divergence free $\Eb^a$ ! This underlines the 
crucial role played by the phases (\ref{alphaabelian}) in the group averaging procedure.

Finaly we rephrase our result above in the language developed in section \ref{sec3B}.
Equation (\ref{UaEabelian})  implies that   $\Ph_{\Eb}=\G$.  $\sym_{\Eb}$ is given by those gauge transformations satisfying  $ e^{-i \int_\Sigma \theta \partial_a \Eb^a }=1$. It then follows that $\Ph_{\Eb}=\sym_{\Eb}$ iff $\partial_a \Eb^a=0$, and the general group averaging sum (\ref{nophasesum}) becomes
\be
\eta^{\G}(|\Eb \ket) = \left\{
\begin{array}{rll}
0 & \text{if }  &\partial_a \Eb^a \neq 0 \\
\bra \Eb| & \text{if } &  \partial_a \Eb^a=0. 
 \end{array} \right.  \label{etacorrect}
\ee

\subsection{Phases in the $SU(2)$ theory} \label{sec4B}

We return to  $SU(2)$ theory in the KS representation. As mentioned at the end of section \ref{sec2}, the states of interest are of the form  $|s,\Eb \ket$ where $s$ is $SU(2)$ gauge invariant. The action of $a=(g,\phi) \in \aut$ on such states (\ref{UasE}) takes the form 
\be
U(a) |s, \Eb \ket = e^{i \alpha(a,\Eb)} |\phi(s), a \cdot \Eb \ket , \quad a = (g,\phi), \label{UasE2}
\ee
so that the spin network label is insensitive to $SU(2)$ local rotations.  States admitting nontrivial rephasings are  those for which there exists $a \in \aut$ leaving the KS labels invariant with $e^{i \alpha(a,\Eb)} \neq 1$. From the discussion of section \ref{sec3B}, such states are  annihilated by the group averaging map.

The nonabelian theory exhibits a general class of triads admitting rephasings in a way that closely resembles the previous abelian example.  Consider a state $|s,\Eb \ket$ with  $\Eb^a$ of rank 1, so that it can be written in the form:
\be
\Eb^a= \nh \, X^a  , \label{r1case}
\ee
where $\nh: \Sigma \to su(2)$ is an $su(2)$ valued scalar with unit norm so that $\tr[\nh \nh]=1$, and
$X^a$ is a unit density weight  vector field.  
It is clear that local $SU(2)$  rotations of the form $g=e^{\theta \nh}$ with an arbitrary function 
$\theta: \Sigma \to \reals$ leave  (\ref{r1case}) invariant, as well as the gauge invariant spin network $s$. The corresponding  phases  (\ref{alpha}) are given by:
\ba
\alpha(e^{\theta \nh},\Eb)  = &\int_\Sigma \tr[ \Eb^a e^{-\theta \nh}\partial_a e^{\theta \nh}] & = \int_\Sigma X^a \tr[ \nh e^{-\theta \nh}\partial_a e^{\theta \nh}] \\
  = & \int_{\Sigma}   X^a \partial_a\theta & = -\int_{\Sigma} \theta \partial_a X^a \label{phaseX},
\ea
where in the third equality we used $\tr[ \nh e^{-\theta \nh}\partial_a e^{\theta \nh}]=\partial_a\theta$, as can be verified for instance  from the expression $e^{\theta \nh}=\cos \tfrac{\theta}{2} \idtwo +2 \sin \tfrac{\theta}{2} \nh$. \footnote{Eq. (\ref{phaseX}) also follows from the general formula (\ref{formulalpha}) discussed in Appendix \ref{phaseat}.}
The situation is then as in the abelian theory, with  $e^{\theta \nh} $ playing the role of local $U(1)$ rotation and $X^a$  playing the role of  abelian electric field. Thus,  if  $\partial_a X^a \neq 0$, there exist non-trivial rephasings and the corresponding state is annihilated by the group averaging map.

Let us finally consider the case of a `pure background' state $|0,\Eb \ket$ with  $\rk(\Eb) \geq 2$.  It is easy to verify that  there are no internal rotations leaving $\Eb^a$ fixed. There could however be  symmetries associated to combinations of diffeomorphisms and local rotations. For this to happen, the  tensor
\be 
\tilde{\tilde{q}}^{ab} := \tr[\Eb^a \Eb^b] , \label{qtt}
\ee 
should admit symmetries. At an infinitesimal level, this corresponds to the existence of vector fields $\xi^a$ satisfying:
\be
\L_{\xi}\tilde{\tilde{q}}^{ab}=0 .\label{lieqtt}
\ee 
In the rank 3 case, condition (\ref{lieqtt}) is not generically satisfied, but only in the special case when the metric admits Killing symmetries. One can similarly show that in the rank 2 case there are no generic solutions to (\ref{lieqtt}), see Appendix \ref{r2sym} for an argument. 

If the  rank $\geq 2$  triad does  admit symmetries, either infinitesimal as in (\ref{lieqtt}) or discrete ones,  one  would then  need to determine the corresponding phases. Recall from the  last paragraph of section \ref{ggs} that such phases give rise to a homomorphism $\S_{\Eb} \to U(1)$. Thus, a necessary condition for the existence of non trivial phases is that the group $\S_{\Eb}$ admits non-trivial homomorphism to $U(1)$. 

The possibility of rank 3 triads admitting symmetries with non-trivial phases is an intriguing one, as it would imply the corresponding states are annihilated by the group averaging map. This would be in striking contrast with the analogous quantization in metric variables \cite{aa-je}, where  \emph{all} metrics yield nontrivial  $\diff$-invariant states. It is however unknown to us whether there actually exist rank 3 triads admitting nontrivial rephasings. In Appendix \ref{commentr3} we discuss further situations where the phase can be shown to be vanishing.

\section{Group averaging: The example of standard LQG} \label{gavediff} \label{sec5} 

In this section we construct  the $\diff$ group averaging map of $SU(2)$ gauge invariant spin networks  
\cite{almmt,alrev} through an application of the strategy described
in  section \ref{sec3}. Our considerations in the well understood context of LQG in this section serve as a preview 
of similar considerations in the context of the  KS representation in the following sections.
As mentioned in section \ref{sec1}, standard LQG is the $\Eb=0$ sector of the KS representation so that the kinematic states
$|s, \Eb=0\rangle \equiv |s\rangle$ are labelled by $SU(2)$ invariant spin networks. 
The automorphism group $\aut$ then 
reduces to the spatial diffeomorphism
group $\diff$.  
The sum over states form of the (putative) group averaging map (\ref{finalsum}) applied to the
spin net state $|s\rangle$ is
\be
\eta(|s \ket)= \eta_{[s]}\displaystyle\sum_{a \in \aut_{ {\rm Orb}(s )}} (U(a)|s\ket )^{\dagger}  .
\label{finalsums}
\ee
We move to a less cumbersome notation tuned to this standard LQG context as follows. 
As indicated above we have that $\aut =\diff$ so that 
$\sym_s:=\{ \phi \in \diff  :  \phi(s)=s \}$ is its symmetry group. Let 
 $\diff/\sym_s$  be the set of right  cosets of  $\diff$ by $\sym_s$. This set is just 
 the set of distinct $\sym_s$-orbits $\phi \,  \sym_s \subset \diff$ for all $\phi \in \diff$. 
The group averaging map (\ref{finalsums}) may then be written as
\be
\eta(|s \ket) :=  \eta_{[s]} \sum_{c \in \diff/\sym_s}   (U(\phi_c)|s\ket)^\dagger . \label{etas}
\ee
Here $\phi_c$ is a choice of 
representative diffeomorphism on each orbit $c$ and we remind the reader that 
$\eta_{[s]}$ is a yet to be determined positive number 
obeying  $\eta_{[\phi(s)]}=\eta_{[s]} \; \forall \phi \in \diff$.

The next step in our strategy is to identify the superselection sector containing $|s\rangle$. This 
is done as follows \cite{almmt,alrev}. Let ${\cal D}$ of \ref{sec3A} be the finite span of spin net states
and so $ O:{\cal D}\rightarrow {\cal D}$ of equation (\ref{3.4}) is  a diffeomorphism invariant operator.
Let the coarsest graph underlying a spin net $s$ be denoted by $\gamma (s)$. Let $|s_1\rangle, |s_2\rangle$ be a pair 
of spin net states. Consider the matrix element $\bra s_1 | O | s_2 \ket$. Suppose that there are infinitely
many diffeomorphisms $\{\phi_i\}$ each of which leave $\gamma (s_1)$ invariant but yield a distinct image 
when applied to $\gamma (s_2)$.
From the commutativity property of observables with diffeomorphisms one has,
\ba
\bra s_2 | O | s_1 \ket &=& \bra s_2 |  O U^\dagger(\phi) | s_1 \ket \\
&=& \bra s_2 | U^\dagger(\phi) O  | s_1 \ket. \label{strobsarg}
\ea
Thus, the state $O  | s_1 \ket$ has the same component along spin networks of the form $|\phi(s_2) \ket$. Since there are infinitely many of them, it follows from $O:{\cal D}\rightarrow {\cal D}$ that $\bra s_2 | O | s_1 \ket=0$. 
This implies that $|s_1\rangle, |s_2\rangle$ are superselected at the kinematic level.\footnote{By kinematic level, we mean kinematic states which are mapped to each other exclusively
 by gauge invariant observables,
as opposed to being mapped on to each by gauge invariant observables or diffeomorphisms.}

We show now that if $\gamma (s_1) \neq \gamma (s_2)$, there are infinitely many elements of $\diff$ which 
move one of them, say $\gamma (s_2)$, and keep the other invariant.  Accordingly, let $\gamma (s_1) \neq \gamma (s_2)$. 
Consider the coarsest graph $\gamma (s_1,s_2)$ which underlies both $s_1$ and $s_2$. It follows that there exists 
an edge $e_{s_1,s_2} \in \gamma (s_1,s_2)$ such that $e_{s_1,s_2}$ is contained in some edge $e_2 \in \gamma (s_2)$
and such that ${\rm Int}(e_{s_1,s_2}) \cap \gamma (s_1)= \emptyset$. Since $\gamma (s_1,s_2)$ is the finite union 
of closed semianalytic edges it follows that there exists a small enough open neighbourhood $U_p$ of a point 
$p \in {\rm Int}(e_{s_1,s_2})$ such that $U_p \cap \gamma (s_1) =\emptyset$ and such that 
$U_p \cap \gamma (s_2)$ is a connected subset of $e_2$. 
Next, consider any semianalytic vector field $v_2$ on $\Sigma$ which is transverse to 
$e_2$ in $U_p \cap \gamma (s_2)$. Let $f$ be a semianalytic function compactly supported within $U_p$.
It follows that there are infinitely many diffeomorphisms generated by the the vector field $fv_2$ which 
yield distinct images of $\gamma (s_2)$ but leave $\gamma (s_1)$ invariant.

Thus, we have that $|s_1\rangle, |s_2\rangle$  are in different kinematical superselection sectors unless $\gamma (s_1)$
coincides with $\gamma (s_2)$. Equivalently from equation (\ref{3.12}) it follows that $\eta (|s_1\rangle ),
\eta (|s_2\rangle )$ lie in the same superselection sector only if there exists a diffeomorphism $\phi$
such that $\phi (\gamma (s_2)) = \gamma (s_1)$. Thus the diffeomorphism invariant 
 superselection sector containing $|s\rangle$ is made up of all spinnets $|s^{\prime}\rangle$ such that 
$\gamma (s^{\prime})= \phi(\gamma (s))$ for some diffeomorphism $\phi$.

Having determined the superselection sectors, the final step in our strategy is to determine the ambiguity
coefficients in such sector.  
A useful heuristic idea underlying this determination is that the coefficients $\eta_{[s]}$ in (\ref{etas}) should   be 
proportional to the `size' of the symmetry group $\sym_s$ in order to compensate for the quotient 
space being summed over: ``$\sum_{\diff} = |\sym_s| \sum_{\diff/\sym_s}$''. 
Now, even though there is no sense in  ``$|\sym_s|$'', one could attempt to 
make sense of \emph{relative sizes} of symmetry groups of spinnets that belong to the \emph{same superselection sector}. 
One way in which this could be done is to identify a suitable `reference subgroup' $\sym^{\rm ref}_{s}$ of $\sym_s$
which is the same for all $s$ in the same kinematic level  superslection sector  
and 
such that $|\sym_s/\sym^{\rm ref}_{s}| < \infty$. 

To implement this idea we proceed as follows.
Let
\be
\sym^0_s := \{ \phi \in \diff : \phi(e)=e \quad \forall e \in \gamma(s) \}, \label{sym0s}
\ee
be the  `trivial action group' of diffeomorphisms preserving the oriented edges of the graph $\gamma(s)$ (denoted  $\text{TDiff}_{\gamma(s)}$ in  \cite{alrev}). It is easy to verify that $\sym^0_s$ is a normal subgroup of $\sym_s$. The corresponding quotient group
\be
D_s:= \sym_s/ \sym^0_s, \label{Ds}
\ee
is the group of discrete symmetries of allowed edge permutation of the spin network $s$. It is a finite group and we denote by  $|D_s|$ its   number of elements or `size'. From the discussion around equation (\ref{strobsarg})
and from the finiteness of $D_s$ it follows that 
$\sym^0_s$ can play the role 
of the desired reference group $\sym^{\rm ref}_{s}$. Let us now see in detail that this is indeed what happens.

Let $s_1$ and $s_2$ be two spin networks based on diffeomorphic graphs (for otherwise property 3 trivializes).  We want to impose the condition
\be
\eta(O |s_1 \ket)[|s_2 \ket]=\eta(|s_1\ket )[O^\dagger |s_2 \ket] \label{etaOs} ,
\ee
for all $O \in \O$.   Since $O : \D \to \D$, the vector $O | s_1 \ket$  admits an expansion of the form,
\be
O | s_1 \ket = \sum_{i=1}^n \lambda_i U(\phi_i) |s_2 \ket + |\chi \ket\quad \text{with} \quad  \bra  s_2 | U(\phi) |\chi \ket = 0 \; \forall \; \phi \in \diff \, . \label{Os1}
\ee
The vectors $\lambda_i U(\phi_i) |s_2 \ket$ represent the components of $O | s_1 \ket$ along the the orbit of $|s_2 \ket$ and are  taken to be orthogonal; $|\chi \ket$ encodes the remaining vectors orthogonal to the span of the orbit of $|s_2 \ket$.

We now use (\ref{Os1}) to evaluate  both sides of (\ref{etaOs}). The left hand side becomes
\ba
\eta(O |s_1 \ket)[|s_2 \ket] & = & \sum_{i} \overline{\lambda}_i \eta(U(\phi_i) |s_2 \ket)[|s_2 \ket] \nn \\
&= &\sum_{i} \overline{\lambda}_i \eta( |s_2 \ket)[|s_2 \ket] \nn \\
&=& \eta_{[s_2 ]}\ \sum_{i} \overline{\lambda}_i .\label{etaO1s}
\ea
To evaluate the right hand side, we first rewrite  $\eta(|s_1\ket )$ as a sum over $\sym^0_{s_1}$ cosets as follows.  Considers the auxiliary map defined by
\be
\eta^0(|s \ket) := \eta_{[s]} \displaystyle\sum_{c \in \diff /\sym^0_{s}}   (U(\phi_c) | s \ket)^{\dagger} . \label{det1s}
\ee
It then follows that:
\be
\eta^0(|s \ket) = |D_s|   \eta(|s \ket). \label{det2s}
\ee
Using (\ref{det1s}) and (\ref{det2s}), the right hand side of (\ref{etaOs}) becomes:
\ba
\eta(|s_1\ket )[O^\dagger |s_2 \ket]  & = & |D_{s_1}|^{-1} \eta_{[s_1 ]} \sum_{c \in \diff/\sym^0_{s_1}} \overline{ \bra s_2 |  O U(\phi_c) | s_1 \ket } \nn\\
& = & |D_{s_1}|^{-1} \eta_{[s_1 ]} \sum_{c \in \diff/\sym^0_{s_1}} \overline{ \bra s_2 |  U(\phi_c) O | s_1 \ket } \nn \\
& = & |D_{s_1}|^{-1}\eta_{[s_1]}  \sum_{c \in \diff/\sym^0_{s_1}} \sum_i \overline{ \lambda_i \bra s_2 | U(\phi_c) U(\phi_i) | s_2 \ket } \nn \\
& = & |D_{s_1}|^{-1} \eta_{[s_1 ]} \sum_i \overline{\lambda}_i \sum_{c \in \diff/\sym^0_{s_1}}  \overline{ \bra s_2 | U(\phi_c) U(\phi_i) | s_2 \ket }, \nn \\
& =: & \eta_{[s_1 ]}   \sum_i \overline{\lambda}_i  x_i  \label{lastcomms}
\ea
where 
\ba
x_i & := &  |D_{s_1}|^{-1} \sum_{c \in \diff/\sym^0_{s_1}}  \overline{ \bra \phi | U(\phi_c)  | s^{(i)}_2 \ket },  \label{sumi} \\
s^{(i)}_2 & := &\phi_i(s_2).
\ea
Notice that the interchange in the sums order leading to (\ref{lastcomms}) is valid since only a finite number of terms in the sum over $\diff/\sym^0_{s_1}$ is non-zero. We now focus on evaluating  (\ref{sumi}). We first notice that since $\bra s^{(i)}_2| O |s_1 \ket = \lambda_i \neq 0$ it follows that $\gamma(s^{(i)}_2)=\gamma(s_1)$. In particular $\sym^0_{s^{(i)}_2}=\sym^0_{s_1}$ and the sum is independent of the orbit representative choices $c \to \phi_c$. Let $c_i:= \phi_i^{-1} \sym^0_{s_1}$ be the $\sym^0_{s_1}$ orbit through $\phi_i^{-1}$. Such term gives a contribution of $1$ to the sum in (\ref{sumi}). All other $\sym^0_{s_1}$ orbits  can be obtained as  $\phi_i^{-1} \phi \sym^0_{s_1}$ for appropriate  $\phi \in \diff$. Nonzero contributions will then come from elements $\phi \in \sym_{s^{(i)}_2}$. It then follows that there are $|D_{s^{(i)}_2}|$ of such terms and so we obtain:
\ba
x_i  & = &  |D_{s_1}|^{-1} |D_{s^{(i)}_2}|.
\ea
The key property of $\diff$ invariance of the relative sizes implies  the numbers above are independent of $i$:
\be
|D_{s^{(i)}_2}|= |D_{s_2}|.
\ee
Assuming $O$ is such that  $\sum_i \lambda_i \neq 0$ (for if no such observable exist then the states would be superselected) we  conclude that:
\be
\eta_{[s_1]}/\eta_{[s_2]}=|D_{s_1}|/|D_{s_2}|.
\ee
Thus, in order to satisfy (\ref{etaOs}), within each sector the ambiguity coefficient must be set as
\be
\eta_{[s]}= C |D_{s}| \label{ambcoefs}
\ee
for some constant $C>0$. 

Since the above presentation slightly differs from the traditional one, let us explicitly see how the two  coincide. In the following we take $\gamma \equiv \gamma(s)$. In  \cite{almmt,alrev}, the group of graph symmetries: 
\be
\sym_\gamma := \{ \phi \in \diff : \phi(\gamma) = \gamma \}
\ee
(denoted  $\diff_\gamma$ in \cite{alrev}), is used instead of $\sym_s$, together with  the corresponding discrete group
\be
D_\gamma := \sym_\gamma/\sym^0_\gamma ,
\ee
(denoted  $\text{GS}_\gamma$ in \cite{alrev}), where $\sym^0_\gamma \equiv \sym^0_s$ is given by (\ref{sym0s}). We now use these groups to rewrite the group averaging map (\ref{etas}) (we set $C=1$ in  (\ref{ambcoefs})):
\ba
\eta(|s \ket ) & = &  |D_s| \sum_{c \in \diff/\sym_s}   (U(\phi_c)|s\ket)^\dagger  \nn \\
&= &|D_s| \sum_{c' \in \diff/\sym_\gamma} \sum_{d \in \sym_\gamma/\sym_s} (U(\phi_{c'}\phi_d )|s\ket)^\dagger \nn \\
&= &|D_s| \sum_{c' \in \diff/\sym_\gamma} \sum_{d \in D_\gamma/D_s} (U(\phi_{c'}\phi_d )|s\ket)^\dagger \nn \\
&= & \sum_{c' \in \diff/\sym_\gamma} \sum_{d \in D_\gamma} (U(\phi_{c'}\phi_d )|s\ket)^\dagger  \label{etalqg}
\ea
where $\phi_{c'}$ and $\phi_d$ are representatives of the cosets $c' \in \diff/\sym_\gamma$ and $d \in \sym_\gamma/\sym_s \equiv D_\gamma/D_s$.   Expression (\ref{etalqg}) takes  precisely the form of the group averaging as given in \cite{almmt,alrev}.

We conclude with a few remarks regarding the role of the groups $\sym^0_s$ and  $\sym_s$ (or $\sym_\gamma$). We first  point out that for both the characterization of superselection sectors and the proof of  well definedness of the group averaging, one  {\em does not} require an explicit characterization of  these  groups. Indeed, we are not aware of such characterization and they may well be complicated  groups with infinitely many connected components. Superselection sectors can be identified by a `large enough' subgroup of $\sym_s$ of the type described at the beginning of the section. 
A perusal of our argumentation in this section indicates
that this `large enough' subgroup is
as follows. Note that every edge $e$ with endpoints removed admits an open cover $\{U^e_{\alpha}\}$  
in  $\Sigma$  which does not intersect any other edge. Consider 
semianalytic vector fields which  compactly supported in each 
open set $U^e_{\alpha}$ and which are parallel to the edge tangent wherever their support intersects $e$. Such
vector fields generate the `large enough' subgroup  of $\sym^0_\gamma$ which suffices to 
identify the kinematic and, from equation (\ref{3.12}),  the diffeomorphism invariant, superselection sector 
containing $\vert s\rangle$. As shown above the underlying graphs of
spin nets in a single such kinematic sector are constrained to  coincide. Thus
all but a finite set of data (namely the edge colorings and the vertice intertwiners) are completely constrained.
As a result the maps on the  remaining finite `free data' are easily understood: they are just the finite edge 
permutation groups referred to above.

To summarize:
The identification of a sufficiently
large (infinite dimensional) subset of $\sym^0_\gamma$  suffices to 
(1) isolate generic superselection sectors and (2) constrain 
an `infinite dimensional' part of the label set of states in each such sector.\\
 The
left over `free data' in the label set is then sufficiently restricted so as to be finite. 
In the LQG case, this finiteness translates to that of the $\diff$ invariant numbers $|D_s|$,
which in turn allows a well defined, consistent evaluation of ambiguity coefficients. 

Finally, we note that while the finiteness of $|D_s|$ allows a proof of {\em existence} of a consistent
group averaging map, the {\em explicit}  identification of $D_s$ and it cardinality is still a non-trivial 
exercise, the degree of non-triviality increasing with complexity of the graph $\gamma (s)$ underlying $s$.

\section{Superselection sectors in the KS representation} \label{sec6}

Given a triad $\Eb^a$, we define the following \emph{rank sets}:
\ba
V_0 & := &\{ x \in \Sigma \; : \; \rk(\Eb)=0\} ,\label{V0}\\
V_1 & := &\{ x \in \Sigma \; : \; \rk(\Eb)=1\}, \label{V1}\\
V_2 & := &\{ x \in \Sigma \; : \; \rk(\Eb) \geq 2 \}. \label{V2} 
\ea
As mentioned in the Introduction,  semianalyticity of the triad  ensures these are  `nice' sets in the sense that they can be expressed as a finite union of semianalytic submanifolds. This follows from properties of the zero level sets, and their complements, of semianalytic functions (see  \cite{lost,milman} and Appendix \ref{saapp} for definitions and results on semianalytic category). More in detail, the rank sets can be described by semianalytic functions as follows. Fix an auxiliary  semianalytic metric $\qo_{ab}$ on $\Sigma$ and  let 
\ba 
v_a^i &:=& \epsilon_{ijk} \qo^{-1/2} \eta_{abc} \Eb^b_j \Eb^c_k \\
g&:=&\qo^{ab}v_a^k v_b^k, \label{g}\\
h &:=&\qo^{-1} \qo_{ab}\Eb^{a}_i \Eb^b_i, \label{h}
\ea
where $\qo= \det(\qo_{ab})$.  It is then easy to verify that:
\ba
V_0 &=& \{ h(x)=0\} , \label{defV0} \\
V_1 &=&  \{ h(x) \neq 0\} \cap \{g(x)=0 \} , \label{defV1} \\
V_2 &=& \{g(x) \neq 0 \}\label{defV2} .
\ea
As discussed in Appendix  \ref{saapp}, sets that are defined by semianalytic functions such as (\ref{defV0}), (\ref{defV1}) or (\ref{defV2}) can always be expressed as finite union of semianalytic submanifolds. In particular, there is a notion of dimension of  such  sets, given by the maximum dimension of its submanifold components. 

To appreciate the non-triviality of this property, let us compare with a case where we had  smooth, rather than semianalytic, fields.  Then $V_0$ would be given by  the  zero level set of a smooth function $h$. By continuity we know $V_0$ is a closed set. But  it turns out that  \emph{any} closed set of the manifold can be realized as the zero level set of some smooth function (see for instance  proposition 3.3.6 of \cite{sabook}). Thus, we  cannot a priori guarantee any property on  $V_0$ (other than it being closed). For instance $V_0$ could have  fractal-like structure and thus be very far from a `finite union of submanifolds' as in the semianalytic case. 

We now use the  rank sets to characterize  superselection sectors.  As in the standard LQG case it is convenient to first characterize ``kinematical superselection sectors'', i.e.  sectors generated by states  admitting non-zero matrix elements on observables. As reviewed in section \ref{sec5}, in the standard LQG case the kinematical sectors correspond to  spin networks based on the \emph{same} graph, while  the superselection sectors of the group averaging map are  given by  spin networks based on \emph{diffeomorphic} graphs.

We now show the following kinematical superselection conditions for KS states $|s,\Eb \ket$ and $|s',\Eb' \ket$:
\ba
\bra s, \Eb | O |s',  \Eb' \ket \neq 0 \implies & i)  &\; \Vo_0=\Vo'_0 ,   \quad \Vb_2=\Vb'_2   \label{cond1}\\
&  ii) &\; \dim(V_{n} \cap (\cup_{n' \neq n}  V'_{n'})) <3 , \; n=0,1,2   \label{cond2}  \\
&  iii) & \gamma(s) \cap \Vo_0= \gamma(s') \cap \Vo_0, \label{cond3}
\ea
where $V_n, V'_n$ are the rank sets of $\Eb, \Eb'$  and $\gamma(s), \gamma(s')$ are the graphs of the spin networks. Notice that the set $V_{n} \cap (\cup_{n' \neq n}  V'_{n'})$ featuring in ii) is again determined by  zeros (and their complements) of semianalytic functions. It is thus  composed of finite union of semianalytic submanifolds, and ``dim'' in ii) refers to the maximum dimension of its submanifold components. 

 The logic in showing (\ref{cond1}), (\ref{cond2}), (\ref{cond3}) is the same as in the spin network case: The existence of infinite gauge transformations leaving one of the states fixed and producing infinitely many distinct images on the other implies their matrix elements on observables vanishes, since observables map  a KS state into  \emph{finite} linear combination of  KS states \emph{and} strongly commute with gauge transformations.  Thus in the following we show how if any of the conditions in (\ref{cond1}), (\ref{cond2}) or (\ref{cond3}) fails to be satisfied, there exist infinitely many of such gauge transformations. 

We  first focus on (\ref{cond1}) and (\ref{cond2}) in a `pure background' case with no nontrivial spin networks. 
We start with  (\ref{cond1}).  It will be convenient to consider the following unions of rank sets:
\ba
V_{01} & := & V_{0} \cup V_1= \{g(x) = 0 \} \label{V01},\\
V_{12} & := & V_{1} \cup V_2 = \{h(x) \neq 0 \} \label{V12} ,
\ea 
and similarly for the primed sets associated to $\Eb'$. To show the first equality in (\ref{cond1}),  assume by contradiction that $\Vo_0 \neq \Vo'_0$. Then there exist a point $x$ such that  $x \in \Vo'_0$ and $x \notin \Vo_0$. The latter means that there  is no open neighborhood  of $x$ that is contained in $V_0$. This in turn is equivalent to the statement that for every open neighborhood $U$ of $x$, $U \cap V_{12} \neq \emptyset$. Taking $U$ such that $U \subset \Vo'_0$, and noting that $V_{12}$ is open, we conclude  that $U \cap V_{12} \neq \emptyset$ is an open set. Diffeomorphisms with support inside  such set  will generate symmetries of $\Eb'$ but not of $\Eb$.  The second equality in (\ref{cond1}) is shown similarly, by noting that $\Vb_2=\Sigma \setminus \Vo_{01}$. If by contradiction there exist a point $x$ such that $x \in \Vo'_{01}$ and $x \notin \Vo_{01}$ we can find an open set  $U \subset \Vo'_{01}$ such that  $U \cap V_{2} \neq \emptyset$. Since $V_2$ is open we can again construct infinitely many symmetries of $\Eb'$ that are not symmetries of $\Eb$. 

Let us now show condition (\ref{cond2}).  Assume by contradiction that  $\dim(V_{n} \cap (\cup_{n' \neq n}  V'_{n'}))=3$.  A case by case analysis shows this leads to a contradiction. We use the fact that dimension 3 sets contain open sets.
\bi
\item $n=0$:  There is an open set  $U \subset V_0 \cap (V'_1 \cup V'_2)$.  Diffeos generated by vector fields with support in $U$ generate infinite symmetries of $\Eb$ but not of $\Eb'$
\item $n=1$:  Either $\dim(V_1 \cap V'_0)=3$  or $\dim(V_1 \cap V'_2)=3$ (or both). In the first case there are diffeos with support inside an open set of $V_1 \cap V'_0$ that  are symmetries of $\Eb'$ but not of $\Eb$. In the second case one can find local internal rotations with support inside $U \subset V_1 \cap V'_2$. These  are symmetries of $\Eb$ but not of $\Eb'$.
\item  $n=2$: Either $\dim(V_2 \cap V'_0)=3$  or $\dim(V_2 \cap V'_1)=3$ (or both). In the first case one finds diffeos that are symmetries of $\Eb'$ but not of $\Eb$. In the second case one finds local internal rotations that are symmetries of $\Eb'$ but not of $\Eb$.
\ei
We thus conclude  (\ref{cond2}). In the case where spin networks are presents, the argument above are still valid, provided one works with  small enough open sets so that they do not intersect the spin network graphs.

Finally, condition (\ref{cond3}) follows from a similar argument as the one given in section \ref{sec5} for the standard LQG case. A detailed proof is given in Appendix \ref{secC4}.

The above kinematical superselection rules translate into the following conditions for superselection sectors of the group averaging map:  A necessary condition for two KS states $| s, \Eb \ket$ and $|  \Eb' , s' \ket$ to lie on the same superselection sector is  the existence of a diffeomorphism $\phi$ such that conditions (\ref{cond1}), (\ref{cond2}) and (\ref{cond3}) hold with $\phi(V'_n)$ in place of $V'_n$ and $\phi(s')$ in place of $s'$.

\section{Group averaging in the absence of rank 1 backgrounds}\label{sec7} 

In this section we restrict  attention to a class of superselection sectors  given by  KS states $|s,\Eb \ket$ such that:
\begin{itemize}
\item[a)] The rank 1 set of the background label $\Eb$ is of zero measure.
\item[b)] There are no infinitesimal symmetries on the rank 2 or 3 regions, that is, no  vector fields exist such that  Eq. (\ref{lieqtt}) holds on the open set $V_2$.
\item[c)] $s$ is $SU(2)$ gauge invariant.
\end{itemize}
It is easy to verify that a state that does not satisfy a) and b) will necessarily lie in a different superselection sector: For a state not satisfying  a) this follows from the discussion of section \ref{sec6};  for a state not satisfying  b) this follows from the fact that such a state admits a one parameter family of symmetries that are not symmetries of states obeying b) and hence they are superselected.  We comment on restriction c) in section \ref{sec9B}.

The considerations of section \ref{sec6}  specialized to the  present setting simplify to: 
\ba
\bra \Eb,s| O | \Eb',s' \ket  \neq 0 \implies & i)  &\; \Vo_0=\Vo'_0 ,   \quad \Vb_2=\Vb'_2 \; \quad (\Vo_1=\Vo'_1= \emptyset )  \label{KSr02i}\\
&  ii) & \gamma(s) \cap \Vo_0= \gamma(s') \cap \Vo_0, \label{KSr02ii}
\ea
where $V_n, V'_n$ are the rank sets of $\Eb, \Eb'$  and $\gamma(s), \gamma(s')$ are the graphs of the spin networks. Conditions $i)$ and $ii)$ are the analogue  of the kinematical superselection condition of coincident graphs in the standard LQG case.  It is easy to verify that in the present setting  $\Vo_{01}=\Vo_0$ and so $\Sigma= \Vo_0 \cup \Vb_2$.\footnote{Since    $V_1 \cap \Vo_{01}= V_{12} \cap  \Vo_{01}$ it follows that $V_1 \cap \Vo_{01}$ is an open set. Since $V_1$ is  of zero measure it cannot contain nontrivial open sets and we conclude that $V_1 \cap \Vo_{01}= \emptyset$.  This implies $V_1 \subset \Vb_2$ from which is easy to verify  that  $\Vb_2=\Vb_{12}$ and hence $\Vo_{01}=\Vo_0$.} Thus the sets in (\ref{KSr02i}) cover all the manifold.

We will  denote KS  labels as: $\psi = (s,\Eb)$, $\psi' = (s',\Eb')$, etc; and the action of $\aut$ on these labels as: $a \cdot \psi \equiv (g,\phi) \cdot (s,\Eb) = (\phi(s),a\cdot \Eb)$. The  group averaging sum  takes the form discussed in section \ref{sec3}:
\be
\eta(|\psi \ket)= \left\{
\begin{array}{lll}
\quad \quad 0 & \text{if }  & \sym_{\psi} \subsetneq \Ph_{\psi} \\
& &\\
\eta_{[\psi]}\displaystyle\sum_{c \in \aut/\sym_\psi} (U(a_c) |\psi \ket )^{\dagger}  &\text{if } &\sym_{\psi} = \Ph_{\psi}  ,
\end{array} \right. 
\label{etaks}
\ee
where $\sym_\psi \equiv \sym_{(s,\Eb)} = \{ a \in \aut \; : \;  U(a) | s,\Eb \ket=| s,\Eb \ket  \}=\sym_s \cap \sym_{\Eb} $ and $\Ph_\psi= \sym_s \cap \Ph_{\Eb} \equiv \sym_s \cap \S_{\Eb}$. Since $\aut$ transformations either rephase a KS state or map it into  an orthogonal one, it follows that all states appearing  in the  sum (\ref{etaks}) are  orthogonal so that $\eta(|\psi \ket) $ is a well defined element of $\D'$. Since the presence of rephasings is an $\aut$ invariant notion (see Appendix \ref{ginvalpha}), it follows that (\ref{etaks}) satisfies the first requirement of  group averaging. The satisfaction of the second requirement is also easily verified. It thus remains to study the third requirement and the corresponding determination of ambiguity coefficients $\eta_{[\psi]}$. The requirement reads:
\be
\eta(O^\dagger |\psi_1\ket)[|\psi_2\ket]=\eta(|\psi_1\ket)[O |\psi_2\ket] \quad \forall \, O, \, |\psi_1\ket ,|\psi_2\ket   . \label{prop3ks}
\ee
We focus on the case where $|\psi_1 \ket$ and $|\psi_2\ket$ are in the same superselection sector and  such that they do not admit non-trivial rephasings, for  otherwise (\ref{prop3ks}) trivializes to $0=0$ (see section \ref{sec3C}).

Similar to section \ref{sec5} we expect that the ratio $\eta_{[\psi_2]}/\eta_{[\psi_1]}$ of  ambiguity parameters will be given by the  `relative size' of the groups  $\sym_{\psi_1}$ and $\sym_{\psi_2}$. In order to make sense of such `relative size' in a way that is compatible with  (\ref{prop3ks})   we need, as in the spin network case,   a common subgroup  of (conjugated versions of) $\sym_{\psi_1}$ and $\sym_{\psi_2}$   such that a) the respective quotient spaces are finite and b) the size of these discrete finite spaces is $\aut$ invariant. The natural analogue of the `reference group' used in the spin network case (\ref{sym0s}) is now given by:
\be
\sym^0_\psi \equiv \sym^{0}_{(s,\Eb)}:= \{ a \in \aut : a|_{\Vb_2}= \id , \; a(e)=e \quad \forall \; e \in \gamma(s) \}, \label{sym0KS02}
\ee
where $e$ denotes the edges of $\gamma(s)$, and $a(e) \equiv (g,\phi)(e):=\phi(e)$. It is easy to verify that $\sym^0_\psi$  is a normal subgroup of $\sym_\psi$. We  assume the corresponding quotient group
\be
D_\psi := \sym_\psi/ \sym^0_\psi
\ee
is finite, this finiteness being expected from that of the group of allowed edge permutations of $\gamma(s)$ and of the discrete symmetries of the background triad in $V_2$. Finally, definition (\ref{sym0KS02}) is $\aut$ covariant in the sense that it satisfies: $\sym^0_{a \cdot \psi}= a \sym^0_{\psi} a^{-1}$. This in turn  implies  the $\aut$ invariance  property on the size of the discrete groups:  $|D_{a \cdot \psi}|=|D_{ \psi}|$.

We can now repeat the argument of section \ref{sec5}  by    making the replacements  $s \to \psi, \phi \to a,\diff \to \aut$ as follows. 

 The  fact that  $O$ preserves $\D$ allows us to write $O | \psi_1 \ket$ as a finite linear combination of states:
\be
O | \psi_1 \ket = \sum_{i=1}^n \lambda_i U(a_i) |\psi_2 \ket + |\chi \ket \quad \text{with} \quad  \bra  \psi_2 | U(a) |\chi \ket = 0 \; \forall \; a \in \aut \, , \label{OE1}
\ee
where  $\lambda_i$ are  nonzero complex numbers,  the vectors $U(a_i) |\psi_2 \ket$ are orthonormal and  $|\chi \ket$ is in the orthogonal complement of the orbit of  $|\psi_2 \ket$. The left hand side of (\ref{prop3ks}) becomes
\be
\eta(O |\psi_1 \ket)[|\psi_2 \ket]=\eta_{[\psi_2 ]}\ \sum_{i} \overline{\lambda}_i .\label{etaO1}
\ee
To evaluate the right hand side, we write $\eta(|\psi_1 \ket)$ as a sum over $\sym^0_{\psi_1}$ orbits:
\be
\eta(|\psi_1 \ket) = |D_{\psi_1}|^{-1} \eta_{[\psi_1 ]} \displaystyle\sum_{c \in \aut /\sym^0_{\psi_1}}   (U(a_c) | \psi_1 \ket)^{\dagger} .
\ee
After a few steps as in  (\ref{lastcomms})  we obtain:
\be
\eta(|\psi_1\ket )[O^\dagger |\psi_2 \ket] =|D_{\psi_1}|^{-1} \eta_{[\psi_1 ]} \sum_i \overline{\lambda}_i e^{- i \alpha_i} \sum_{c \in \aut /\sym^0_{\psi_1}}  \overline{ \bra \psi_2 | U(a_c)  | \psi^{(i)}_2 \ket }, \label{lastcomm}
\ee
where $\psi^{(i)}_2 \equiv a_i \cdot \psi_2$ and $\alpha_i = \alpha(a_i,\Eb_2)$ so that $U(a_i) |\psi_2 \ket = e^{i \alpha_i}|\psi^{(i)}_2 \ket$.  Since $\bra \psi^{(i)}_2 | O | \psi_1 \ket = e^{i \alpha_i}\lambda_i \neq 0$ we conclude   $\sym^0_{\psi^{(i)}_2} = \sym^0_{\psi_1}$.  Let $c_i:= a_i^{-1} \sym^0_{\psi_1} \in \aut/\sym^0_{\psi_1}$ be the $\sym^0_{\psi_1}$-orbit along $a_i^{-1}$. Such a term gives a contribution of $e^{i \alpha_i}$ in the second sum of (\ref{lastcomm}). All other orbits  can be obtained as  $a_i^{-1} a \sym^0_{\psi_1}$ for appropriate  $a \in \aut$. Nonzero contributions come from elements $a \in \sym_{\psi^{(i)}_2} $ and so there are  $|D_{\psi^{(i)}_2}|$ terms contributing to the sum. Finally, the $\aut$ invariance of the sizes of the discrete groups imply $|D_{\psi^{(i)}_2}|= |D_{\psi_2}| \;,  i=1,\ldots,n$. Equating the result with (\ref{etaO1}) we obtain: $\eta_{[\psi_2 ]}=|D_{\psi_1}| |D_{\psi_2}|^{-1} \eta_{[|\psi_1]}$. Thus, in order to satisfy (\ref{prop3ks}) for KS states $\psi$ in the superselection sector of interest  we must set:
\be
\eta_{[\psi]}= C |D_{\psi}| \label{etacoefE}
\ee
for some constant $C>0$.

\section{Rank 1 background triads} \label{sec8}
In this section we summarize our (partial) understanding of group averaging related subtleties in the case of background triad labels with rank 1 support sets of nonzero measure.

The organization of the material is as follows.  
In section \ref{sec8A} we discuss  symmetries of triads with rank 1 regions and their associated phases. In section \ref{sec8B} we describe superselection conditions arising from these symmetries and comment on difficulties when  spin networks are present. In section \ref{sec8C} we illustrate  subtleties of implementing the group averaging of section \ref{sec7} in the presence of rank 1 backgrounds.

\subsection{Symmetries of rank 1 triads and their phases} \label{sec8A}
We focus on infinitesimal symmetries, from which the corresponding  phase can be obtained by the following formula that we prove in Appendix \ref{ginvalpha}:
\be
\alpha(e^{ (\Lambda,\xi) \cdot}, \Eb) =\int_{\Sigma} \tr[ \Eb^a \partial_a \Lambda] \; , \quad \text{for } \; (\Lambda,\xi) : \; [\Lambda,\Eb^a]- \L_\xi \Eb^a =0 .\label{formulalpha}
\ee

\subsubsection{Constant rank case} \label{cr1}
Recall from section \ref{sec4B} that rank 1 triads: 
\be
\Eb^a = \nh X^a \label{r1}
\ee 
average to zero if $\partial_a X^a \neq 0$. Thus we  restrict attention to `divergence free' rank 1 triads, i.e. $\partial_a X^a=0$. In this first part we focus in the case where (\ref{r1}) holds everywhere  on $\Sigma$ with $X^a \neq 0$; in the next subsection we discuss the case where (\ref{r1}) holds on some region of $\Sigma$ of nonzero measure.    

Symmetries  of (\ref{r1}) are given by $a=(g,\phi) \in \aut$ satisfying
\be
(g,\phi) \cdot (\nh \, X^a) = g \, (\phi_* \nh)\, g^{-1} \, \phi_* X^a = \nh \, X^a. \label{symrank1}
\ee
Since `internal' and `external' indices are factorized,  the symmetry condition (\ref{symrank1}) translates into the two equations: $\phi_* X^a =  X^a$ and  $g \phi_* \nh g^{-1} \,  =\nh$. At the infinitesimal level, they correspond to the following conditions:
\ba
 0&=& \L_{\xi} X^a \label{phiX}  ,\label{xiX}\\
 0&=& [ \Lambda , \nh ]  -\L_{\xi} \nh    .\label{phinh}
\ea
The most general solution of (\ref{phinh})  is given by,
\be
\Lambda=-[\L_\xi \nh,\nh] + \theta \nh ,\label{lamxi1}
\ee
for  any function $\theta : \Sigma \to \reals$. When $\xi^a=0$, (\ref{lamxi1}) reduces to the
local rotations about $\nh$ discussed in section \ref{sec4B}. Nonzero vector fields $\xi^a$  satisfying (\ref{phiX}) give rise to additional symmetries we now describe.

Let us first rewrite  equation (\ref{phiX}) in a way that facilities the discussion of its solutions and corresponding phases. 
Denote the two-form  dual to $X$ by $\w$ so that $\w_{ab}:=\eta_{abc}X^c$. 
The divergence-free property of $X^a$ translates into the condition $d \w=0$. From the Lie derivative formula: 
$\L_\xi \w= (d i_\xi + i_\xi d) \w$,  it follows that  (\ref{phiX}) is equivalent to  
$d i_\xi \w=0$. Let us now restrict attention to the case of simply connected  $\Sigma$ so that we can rewrite this condition as 
 $i_\xi \w =d f$ for some function $f$. To summarize: for divergence free vector fields $X^a$, we have that\ba
\L_{\xi} X^a =0 &\iff&  i_\xi \w =d f \nn \quad (\text{case of simply connected }  \Sigma) \\
&\iff &X^{[a}\xi^{b]}= -\frac{1}{2}\eta^{abc}\partial_c f \quad \text{for some } f : \Sigma \to \reals ,\label{Xxi}
\ea
where the third equation reformulates the second equation in terms of the original vector field $X^a$.
The simplest class of solutions to (\ref{Xxi}) are given by taking $f=$ constant and $\xi^a$ of the form $\xi^a= g X^a$ for \emph{any} density weight $-1$ scalar $g$.  There may also be    `transversal' symmetries with $\partial_a f \neq 0$.  The issue of fully characterizing these additional transversal symmetries remains an open problem to us which we comment on  below. Fortunately it is still possible to show the phases vanish without explicit knowledge of the solutions to (\ref{Xxi}): Using  $(\ref{formulalpha})$ with  $(\Lambda,\xi)$ satisfying  (\ref{lamxi1}) and (\ref{Xxi}) we find,
\ba
\alpha(e^{(\Lambda,\xi) \cdot},\nh X) & =& \int_{\Sigma} X^a \tr[\nh \partial_a\Lambda] = -\int_{\Sigma} X^a \tr[ \partial_a \nh  \Lambda  ]  \nn\\
& =& \int_{\Sigma}  X^a \xi^b \tr[\partial_a \nh [\partial_b \nh,\nh]  ] = \int_{\Sigma}  X^a \xi^b \tr[[\partial_a \nh,\partial_b \nh]\nh  ] \nn \\
& =& -\int_{\Sigma} \eta^{abc}\partial_c f \tr[\partial_a \nh \, \partial_b \nh \, \nh  ]  = \int_{\Sigma} \eta^{abc} f \tr[\partial_a \nh \, \partial_b \nh \, \partial_c \nh  ] =0. \label{alphar1c}
\ea
The  vanishing follows from the fact that since $n_1^2+n_2^2+n_3^2=1$, the  differentials $\partial_a n_i\; , i=1,2,3$ are not linearly independent, and thus their antisymmetric product vanishes.

We conclude the section by summarizing our  understanding on the solutions of (\ref{Xxi}) and hence on the continuous symmetries of rank 1 triads.  Some additional details  are given in Appendix \ref{symr1}. 

We start by describing a class of configurations for which we do have  a complete description of  $\xi^a$ and $f$ satisfying (\ref{Xxi}). Since $\w$ is closed and nowhere vanishing, it can be thought of as a presymplectic form on $\Sigma$. Proceeding in analogy with the description of phase spaces with gauge symmetries (see for instance Appendix B of \cite{aabook}), consider the `reduced phase space'  $\bar{\Sigma}:=\Sigma/\sim$ where $x \sim y$ iff they lie along the same orbit of $X^a$. Let us now restrict to the case where $X^a$ is such that  $\bar{\Sigma}$ is a (two dimensional) manifold. In this case, $\w_{ab}$ can be realized as the pull-back under the projection map $\pi:\Sigma \to \bar{\Sigma}$ of a closed two form $\bar{\w}_{ab}$ on $\bar{\Sigma}$. Solutions to (\ref{Xxi}) can then be described as follows: Take any function $\bar{f}$ on $\bar{\Sigma}$. Let $\bar{\xi}^a= \bar{\w}^{ab}\partial_b \bar{f}$ be the corresponding `Hamiltonian' vector field on $\bar{\Sigma}$. Such vector field defines a (non-unique) vector field $\xi^a$ on $\Sigma$ such that $\pi_* \xi^a= \bar{\xi}^a$. The ambiguity in the definition of $\xi^a$ is given by vector fields parallel to $X^a$.  It is then easy to verify  that i) the vector field $\xi^a$ obtained in this way solves (\ref{Xxi}) with  $f=\bar{f} \circ \pi$ and ii) any solution to (\ref{Xxi}) can be seen as arising from this construction. In this manner we  obtain an  explicit description of the solution to (\ref{Xxi}).

In a  general case, where $\bar{\Sigma}$ may not be a manifold we have less control over the `transversal' symmetries. Even though there exists a simple \emph{local} characterization of solutions to (\ref{Xxi})  (see Appendix \ref{symr1}), we do not know if these local solutions extend to  \emph{global} solutions of (\ref{Xxi}).

\subsubsection{Varying rank  case} \label{genconf}
We now focus in the case of generic triads, by which we mean triads of variable rank that do not have  infinitesimal symmetries associated to the rank $\geq 2$ regions. In this case the phases associated to infinitesimal symmetries of the triad are determined by the rank 1 region:
\be
\alpha(e^{(\Lambda,\xi) \cdot},\Eb) =\int_\Sigma \Eb^a \partial_a\Lambda = \sum_{n=0}^{2} \int_{V_n} \Eb^a \partial_a\Lambda =\int_{V_1} \Eb^a \partial_a\Lambda, \label{intu1}
\ee
since  $\Eb^a|_{V_0}=0$ and  $\Lambda|_{V_2}=0$.  From the analysis of the constant rank 1 case, it is immediate to conclude that
\be
[\Lambda,\Eb^a]- \L_\xi \Eb^a =0   \implies \Lambda|_{\Vo_1}= -[\L_\xi \nh,\nh] + \theta \nh , \quad \L_\xi X^a|_{\Vo_1}=0, \label{lamv1}
\ee
for some function $\theta$ on $\Vo_1$. Let us now restrict attention to the case where $\Vo_{01}$ is simply connected so that the second condition can be written as in (\ref{Xxi}):
\be
\partial_a f = -\eta_{abc}X^b \xi^c \quad \text{on } \; \Vo_{01}\label{partialf}
\ee
for some function $f$. In (\ref{partialf})  we have  extended   $X^a$ to $\Vo_{01}$ by setting $X^a|_{V_0}:=0$.  In order to extend the computation  (\ref{alphar1c}) to the present case, we need to take into account  boundary terms.\footnote{We assume the set $V_1$ is such that we can apply Stokes theorem. It is likely that the rank sets are nice enough so that Stokes theorem applies, but we have not studied this issue in detail.} There are two integration by parts in (\ref{alphar1c}), occurring in the second and sixth equalities. These yield the following contributions:
\ba
\int_{V_1} \Eb^a \partial_a\Lambda = \int_{\partial V_1} dS_a X^a \tr[\nh  \Lambda] -\int_{\partial V_1} dS_a \eta^{abc}f \tr[\partial_b \nh \, \partial_c \nh \, \nh  ], \label{alphabdy}
\ea
for $\Lambda$ as in (\ref{lamv1}) and $f$ as in (\ref{partialf}). The first integral is given by 
\be
\int_{\partial V_1} dS_a X^a \tr[\nh  \Lambda] =\int_{\partial V_1} dS_a X^a \theta. \label{firstint}
\ee
Since  $X^a|_{\partial V_1 \cap \partial V_0}=0$ and $\theta|_{\partial V_1 \cap \partial V_2}=0$ (by continuity of $\Eb^a$ and $\Lambda$  respectively), it follows that (\ref{firstint}) vanishes.  The second integral requires lengthier discussion. To keep notation simple  we now assume that $V_1$ is connected, the generalization to multiple components being straightforward.  To begin with we notice that since  $X^a|_{\partial V_1 \cap \partial V_0}=0$ and $\xi^a|_{\partial V_1 \cap \partial V_2}=0$, equation  (\ref{partialf}) implies  $\partial_a f |_{\partial V_1}=0$. In particular,  $f$ takes a constant value on  each connected component of the boundary of $V_1$.
 Denoting by  $\partial V_{1}^{(k)}, k=1, \ldots, N$ the connected components of $\partial V_1$ and by $f_k=f|_{\partial V_{1}^{(k)}}$ the constant value taken by $f$ on each component,\footnote{$f$ is defined up to an additive  constant, and so are the $f_k$'s. The total integral (\ref{intu1sum}) however, is independent of this constant, as it should be. This is explicitly seen in expression  (\ref{finalalpha}).} the integral can be written as:
 \be
 \int_{\partial V_1} dS_a \eta^{abc}f \tr[\partial_b \nh \, \partial_c \nh \, \nh  ]= \frac{1}{2}\sum_{k=1}^{N} f_k \int_{\partial V_{1}^{(k)}} \tr[\nh d \nh \wedge d \nh  ]. \label{intu1sum}
 \ee
We notice that the two-form being integrated on the right hand side is the pull-back of the area form on the two-sphere under the map $\nh: \partial V_{1}^{(k)} \to S^2 \subset su(2)$.  If $m_k \in \Z$ is the degree of such map\footnote{The numbers $m_k$ are topological and in particular $\aut$-invariant. The invariance under gauge transformations can be seen explicitly from the identity $\tr[\nh_g d \nh_g \wedge d \nh_g  ]-\tr[\nh d \nh \wedge d \nh  ] = d(\tr[g^{-1}d g \nh ])$  where $\nh_g=g \nh g^{-1} $.}, the integral is  given by:
\be
\int_{\partial V_{1}^{(k)}} \tr[\nh d \nh \wedge d \nh  ] = 4 \pi m_k \;, \quad \sum_{k=1}^N m_k=0 .\label{mk}
\ee
The condition that the degrees add  to zero follows from Stokes theorem and the fact that $\tr[\nh d \nh \wedge d \nh  ]$ is closed (see  paragraph following (\ref{alphar1c})). In particular, if $\partial V_1$ is connected we can ensure the phases vanish.

To determine the fate of the phases in more general cases one needs to study the possible values $f_k$ can take. This in turn is related to the problem of understanding the  `transversal' symmetries of $X^a$. As  mentioned at the end of \ref{cr1},  this remains  an open problem to us  and thus the issue of the phases (in the case of $V_1$ with multiple boundaries) remains unsettled.   

To summarize, the phase associated to a infinitesimal symmetries $(\Lambda,\xi)$  takes the form (case of connected $\Vo_1$ and simply connected $\Vo_{01}$):
\be
\alpha(e^{(\Lambda,\xi) \cdot},\Eb)= -2 \pi \sum_{k=1}^N f_k m_k \quad \text{with} \quad \sum_{k=1}^N m_k=0, \label{finalalpha}
\ee
where the sum is over all connected components of $\partial V_1$, with $f_k$ the value of $f$ at each component and  $m_k$ the degree of $\nh$ on each component.

\subsection{Superselection conditions in the presence of rank 1 triads} \label{sec8B}
The symmetries discussed in the previous section imply new superselection conditions in addition to the ones described in section \ref{sec6}. Let us first focus on the case of `pure background' states. Accordingly consider two states $| \Eb \ket$ and $|\Eb' \ket$  with nonzero  overlap on some observable. The condition $\Vb_2= \Vb'_2$ (\ref{cond1}) implies $\Vo_{01}=\Vo'_{01}$ since $\Vo_{01}=\Sigma \setminus \Vb_2$.  Diffeomorphisms generated by  vector fields parallel to  $X^a$ and $X'^a$ can be used to conclude (see Appendix \ref{appcond3p}):
\be
\bra \Eb| O |\Eb' \ket \neq 0 \implies \; X^{[a}X'^{b]}=0 \quad \text{on }\;  \Vo_{01}= \Vo'_{01} ,\label{cond3p}
\ee
(recall the densitized vector field  $X^a$ on $\Vo_{01}$ is the result of extending $X^a$ to $V_0$ by setting $X^a|_{V_0}=0$).
The above condition is very general but at the same time not the strongest possible:  In all  configurations we can envisage,   there are always enough `transversal' symmetries of the vector fields $X^a$ that allows one to conclude the stronger condition: $\bra \Eb| O |\Eb' \ket \neq 0 \implies X^a = c X'^a \quad \text{on } \; \Vo_{01}$ for some constant $c$.   However, as  in the constant rank case, we have not been able to construct a proof (or counterexample) of this stronger condition.

Let us now consider the inclusion of spin networks. Thus we now have $|\psi \ket= |s,\Eb \ket$ and $|\psi' \ket= |s',\Eb' \ket$ with nonzero overlap on an observable.   What can we say about the portions of the spin network graphs lying in $\Vo_{01}=\Vo'_{01}$?   From the standard LQG argument we know that edges of the spin networks $s$ and $s'$ contained on  $\Vo_0=\Vo'_0$ must agree.

The first observation is that edges $e$ that are entirely contained in $\Vo_1 \cap \Vo_1'$ must also agree, since otherwise one can use diffeomorphisms generated by  vector fields parallel to $X^a$ to produce infinitely many symmetries of one state that change the other. Since edges on  $\Vo_0$ must also also agree, it remains to describe the situation for edges intersecting 'transition' regions between rank 1 and rank 0.

There are many different realizations of such a situation. While case by case studies of various examples seem tractable, in the absence of exhaustive results, we conclude our discussion of superselection here.

%Since it is possible to envisage many different such  situations we will simply mention two cases, one for which there are restrictions and one for which there are not.

%Consider first the case of a two dimensional boundary  between rank 1 and rank 0 regions. The divergence free property of $X^a$ implies that $X^a$   becomes parallel to the boundary as it approaches it, up to lower dimensional sets where it may be  approaching transversally. On  the regions  where $X^a$ is parallel to the boundary one can find  symmetries of the triads given by diffeomorphisms generated by vector fields parallel to $X^a$ that are nonzero at the boundary. Such symmetries can be used to conclude that edges either contained or intersecting this boundary regions must coincide. Thus for such regions we obtain the same conditions as in $\Vo_0$ and $\Vo_1$.

%As the  second example consider a case where  $X^a$ has a flow line that start and ends at two points where $X^a$ vanishes (since $X^a$ is divergence free, such lines can at most span a two dimensional set). Consider the case where such line is isolated, so that it is an intrinsic property of $X^a$ and hence it is preserved by the symmetries of $X^a$.   The inclusion of an edge that coincides with such line (including the end points) does not change the symmetries of the state. Thus one of the states could have such edge and not the other. 

\subsection{Subtleties  in the  group averaging} \label{sec8C}
 A key ingredient in showing well definedness of the group averaging of  section  \ref{sec7} is the existence of a  `reference group' $\sym^0_\psi$ (\ref{sym0KS02}) with respect to which we  could find the relative size of the symmetry groups $\sym_\psi$.  In the presence of rank 1 regions however, we do not know of a general prescription to  define the appropriate reference group. 

To illustrate the difficulties let us  focus on  pure background states $|\Eb \ket \equiv |0,\Eb \ket$ with  $\Eb \neq 0$. We would like to  find a general definition of `reference group' $\sym^0_{\Eb} \subset \sym_{\Eb}$ such that: a) the quotient $\sym_{\Eb}/\sym^0_{\Eb}$ is always finite; b)    $\sym^0_{\Eb'}= \sym^0_{\Eb}$ if $|\Eb \ket$ and $|\Eb' \ket$ are in the same kinematical superselection sector; and c) $a \sym^0_{\Eb}a^{-1}=\sym^0_{a \cdot \Eb}$.

Since we want  $\sym_{\Eb}/\sym^0_{\Eb}$  to be composed of discrete transformations, a natural candidate for $\sym^0_{\Eb}$ is given by the continuous group generated by infinitesimal symmetries:
\be
\sym^0_{\Eb} = \{ e^{(\Lambda_1,\xi_1)\cdot} \ldots e^{(\Lambda_n,\xi_n)\cdot} \in \aut : [\Lambda_i,\Eb^a]- \L_{\xi_i} \Eb^a =0 , \; \sint \tr[\Eb^a \partial_a \Lambda_i]=0  \}. \label{sym00}
\ee
It is easy to verify that $\sym^0_{\Eb}$ is a normal subgroup of $\sym_{\Eb}$.  Unfortunately, as we show below,  it is not necessary that $\sym_{\Eb}/\sym^0_{\Eb}$ is finite. Thus, although it is conceivable that (\ref{sym00}) may serve as a reference group for certain superselection sectors, it cannot be used as a general prescription. 

We now give an example where $\sym_{\Eb}/\sym^0_{\Eb}$ is of infinite size. The possibility  of  `infinite discrete symmetries'  arises from submanifolds of $\Sigma$ whose  mapping class group \cite{mcg} is of infinite size.\footnote{Similar `infinite discrete symmetries' could be encountered in the  spin network case of section \ref{sec5} if one (wrongly) uses as a reference group the subgroup of (\ref{sym0s}) given by diffeomorphisms of the type $\phi = e^{\L_{\xi_1}} \ldots e^{\L_{\xi_n}} $ with $e^{\L_{\xi_i}}(e)=e$ for all edges $e$ of the spin network. The difficulty in the present case is that we do not know what the analogue of $\sym^0_s$ (\ref{sym0s}) for general `pure background' states is, let alone  for general KS states. \label{fncs}} We first give a two dimensional  example and then show its three dimensional counterpart. Let   $\Sigma= \reals^2$ with polar coordinates $(r,\theta)$ and consider an $su(2)$ electric field  $\Eb^a$  such that: a) $V_1=\{1 \leq r \leq 2 \}$ ,   $\Eb|_{V_1}= \t_3 \frac{\partial}{\partial \theta}$; b) $V_2= \Sigma \setminus V_1$ and $\Eb|_{V_2}$ does not admit infinitesimal symmetries. 

Consider a vector field $\xi^a= f(r) \frac{\partial}{\partial \theta} $ where $f(r)$ is semianalytic and satisfies $f(r\geq 2)=0$ and $f(r \leq 1)=2 \pi$. Let $\phi_t= e^{t\L_\xi}$ be the one parameter group of diffeomorphisms generated by $\xi$. Then  $\phi_t$ generates symmetries of $\Eb|_{V_1}$ for all $t$. These however do not represent one parameter group of symmetries of $\Eb$ since by assumption $\Eb|_{V_2}$ does not admit infinitesimal symmetries. However, for the values $t= n \in \Z$ they do yield symmetries of $\Eb$ since  $\phi_{t=n}|_{V_2}= \id$. Now, the diffeomorphisms $\phi_n|_{V_1}$ provide representatives of the so called mapping class group of the annulus,  $\text{MCG}(V_1)=\Z$  \cite{mcg}, and so they yield different elements of the  (diffeomorphism part) of the quotient group $\sym_{\Eb}/\sym^0_{\Eb}$, which implies it is of infinite size.

A three dimensional version of the example above exhibiting an infinite quotient group can be constructed as follows.  Let $\Sigma= \reals^3$ with cylindrical coordinates $(z,\rho,\varphi)$ and take $\Eb$ such that  $V_1=\{1 \leq (\rho-a)^2+z^2 \leq 4 \}$ with $a$ large enough so that $\{ (\rho-a)^2 +z^2=4\}$ is a well defined torus. Let $\Eb|_{V_1}=\t_3 [(\rho-a) \partial_z-z \partial_\rho]$ and $\Eb|_{\Sigma\setminus V_1}$ of rank $\geq 2$ admitting no infinitesimal symmetries.  Thus a cross section  $\varphi=$constant corresponds to  the type of configuration  of the two dimensional example. The $t=n$ diffeomorphisms generated by $\xi^a=f(\sqrt{(\rho-a)^2+z^2})[(\rho-a) \partial_z-z \partial_\rho]$ with $f(r)$ as before  correspond again to different elements in  $\sym_{\Eb}/\sym^0_{\Eb}$ and so the quotient group is of infinite size.

We emphasize that in the example above it \emph{is} possible to identify the superselection sector as well as an appropriate reference group: By arguments similar of the type discussed in sections \ref{sec6} and \ref{sec8B} one can show the kinematical superselection sector is given by states $|\Eb' \ket$  such that: $\Vb'_2=\Vb_2, V'_1=V_1$, $\Eb'|_{V_2}$ does not admit infinitesimal symmetries and $\Eb'|_{V_1}=c \Eb|_{V_1}$ for some constant $c \neq 0$. The reference group can  then be  taken to be $\sym^{\rm ref}_{\Eb'}=\{a \in \sym_{\Eb'}:a|_{V_2}=\id \}$. The  quotients $\sym_{\Eb'}/\sym^{\rm ref}_{\Eb'}$ are then finite, corresponding to discrete symmetries of $\Eb'|_{V_2}$. It is however  not clear that this prescription will work in other sectors. 

\section{Discussion} \label{sec9}
\subsection{Summary and discussion of results.} \label{sec9A}

The KS representation is based on a Hamiltonian formulation of gravity in terms of $SU(2)$ connections and triads.
The group of gauge transformations 
generated by the $SU(2)$ Gauss Law and spatial diffeomorphism constraints is referred to
in this work as the automorphism group Aut. As shown here,  
the correct unitary action of Aut in the KS representation
involves hitherto unnoticed phase contributions. 
In this paper we incorporated these phase contributions into an analysis of the imposition of  invariance 
under the group Aut in the KS representation via group averaging techniques. 

Since we do not know (and are not aware) of any well defined group invariant measure on $\aut$, 
our starting point was to define the action of  the putative group averaging map on a state
 to be a sum over gauge related states. The strategy was then to check if this sum over states leads to
a well defined map which satisfies the properties of a group averaging map (see section \ref{sec3A}). 
We showed that  if there exist automorphisms which nontrivially rephase a state then such 
a state is in the kernel of this  
sum over states map. Since any group averaging map must have this property 
(see section \ref{sec3B}), the putative sum- over
-states group averaging map passes this test. We also showed, as required for the consistency of the group averaging map,
that states in this `rephasing' related  kernel of the sum- over- states map lie in superselection sectors 
distinct from the ones which contain
states which do not admit such rephasings. 

Next, we showed that nontrivial rephasings arise generically for KS states labelled by 
background  triads whose rank 1 support sets are of non- zero measure. We also showed that such rephasings can occur,
in principle, for states labelled by non- degenerate background triads with appropriate symmetries.

Next, we derived the main result of this paper, namely the (partial) labelling of superselection sectors
by the (diffeomorphism equivalence classes of) rank support sets (upto sets of zero measure) 
of the background triad label of the KS spin net being averaged. 

Finally, we showed, modulo the assumptions of section \ref{sec9C} that the group averaging map is well defined
for the superselection sectors in which the rank 1 support set of the background triad label is of zero measure.
We remind the reader that for generic states in  this sector, there are no rephasing subtelities.

The open questions, therefore, all pertain to the case of rank 1 support sets of non- zero measure.
Due to the rich structure of the symmetries of such triads, we were unable to derive results of wide
generality. However, we did derive a number of case by case results which we presented in section \ref{sec8}.
Notwithstanding our expectation of the complicated structure of the symmetry group, we do expect that it is most likely 
that further progress does 
{\em not}  entail a detailed understanding of 
the symmetry group of the  labels of the KS state being averaged.
 Rather, we expect that  
the identification of a sufficiently
large (infinite dimensional) subset of this group suffices to 
(1) isolate generic superselection sectors and (2) constrain 
an `infinite dimensional' part of the label set of states in each such sector. If
the left over `free data' in the label set is sufficiently restricted so as to be finite, it should be 
possible to demonstrate a consistent evaluation of ambiguity coefficients. Indeed, as we emphasized in section \ref{sec5},
this is exactly
what happens in LQG wherein  a sufficiently large subset of 
the symmetry group constrains the graphs of superselected states to coincide (up to diffeomorphisms)
and the left over `free data' of a state $|s\rangle$ can be mapped to that of another by elements of the  finite 
graph symmetry group $D_{\gamma}$. From this perspective (see section \ref{sec5} for a detailed discussion)
 the two ingredients needed to complete the group averaging 
procedure are (i) the identification of a sufficiently large set of symmetries which constrain the values of 
all but a finite set of state labels for states within a superselection sector and (ii) a sufficiently large
`reference' subgroup of symmetries underlying all states in a superselection sector where `large enough'
implies that group of all symmetries of the label set for any state in this sector modulo the 
reference symmetries is finite.

\subsection{On the restriction to gauge invariant spin nets.} \label{sec9B}

In this section we comment on our restriction to gauge invariant spin net labels and discuss
ways to generalise our group averaging procedure to the gauge variant case. 
We provide only rough arguments which we hope will serve as guidlines for future attempts at an exhuastive treatment.

As in Sahlmann's work \cite{hs}, we have restricted attention to KS states with gauge invariant spin net labels.
The reason to do so stems from the desire to avoid the following complication,  related to the averaging over 
the `internal gauge transformation' part of $\aut$, which arises when this restriction is relaxed. 
Consider a KS spin net  $|s, \Eb\ket$ with $s$ being an $SU(2)$ gauge {\em variant} spin net and 
consider its group average over $SU(2)$ transformations 
defined through a {\em sum} of the type (\ref{orbit}) over its $SU(2)$ gauge 
related images. 
Suppose there are infinitely many internal  $SU(2)$ rotations which leave the label $\Eb$ invariant but change
the label $s$. The action of these transformations on the state $|s, \Eb\ket$ yields infinitely many 
states which are {\em not} orthogonal because of the non- orthogonality of such $SU(2)$ gauge related 
states in standard LQG. As a result, it looks as if the action of  the `sum over bras'- group averaged state  
on, say, the 
state $|s, \Eb\ket$ is ill defined. In other words, it seems as if this sum does not reside in $\D'$ so that the
putative averaging map violates (\ref{prop1}).
If, on the other hand, the action of these transformations is {\em integrated},
as in standard LQG,
over an appropriate measure deriving from the Haar measure on $SU(2)$, we expect
the resulting group averaged state to  reside in $\D'$. However, such a procedure would require the choice
of group averaging `measure' to be further tailored to the nature of the orbit of the state being averaged.
A simple way to avoid this complication is to restrict the spin net labels to be gauge invariant.

The consequence of this restriction is that 
one is then obliged to restrict attention to  gauge invariant observables which preserve the finite span of KS states
with gauge invariant spin net labels. This is what we have done in this paper. However,
as far as we can see, there is no reason for  gauge invariant observables to
map such states exclusively into themselves 
i.e. it is conceivable that  there exist gauge invariant observables which map KS states with
 $SU(2)$ gauge invariant spin net labels to ones with  $SU(2)$  gauge variant spin net labels. Therefore this
restriction seems to  entail a loss of generality and one would like to remove it.
%Before suggesting a procedure which handles the averaging of KS states which are not subject to this restriction, 
%it is pertinent to re- evaluate the argument discussed above for enforcing the restriction.
It is pertinent to note that in the case of generic triads of rank 2,3 there are {\em no} internal rotations which 
preserve the triad so that the argument of the previous paragraph cannot be applied to this case
\footnote{This was not realised in the pioneering work of Reference \cite{hs}. }.
However,  in the case of rank 0 and rank 1 triads, clearly there are infinitely many internal rotations  
which preserve the  background triad. 
Let us focus on the rank 0 case first. Accordingly, 
consider the averaging of $|s, {\bar E}\ket$ with ${\bar E}$ such that   $V_0$ is of non- zero measure.
Further, let  $s$ be a gauge variant spin net which is an element of the basis of extended spin networks 
contructed in Reference \cite{area}.
%Recall that a  basis of gauge variant spin nets 
%was constructed in Reference \cite{area}. 
Recall that each such basis  element 
has, in addition to the edge labels and invariant intertwiners at vertices, an assignation of vectors, $\nu_v$
in appropriate representations, one for each vertex $v$. For the subspace of gauge invariant spin nets, these
vectors are all in the trivial $j=0$ representation. 
Let us refer to vertices $v$ labelled by $j_v\neq 0$ vectors
as gauge variant and the remaining ones as gauge invariant vertices. Given any $g(x)\in {\cal G}$, such 
basis states transform by a rotation of the vector labels $\nu_v\rightarrow g_{j_v}(v)\nu_v$ where $g_{j_v}(v)$ is
the matrix representative of $g(x=v)$ in the spin $j_v$ representation. 
Let $\mathring{V_0}$ contain a 
gauge variant vertex $v_0$. Clearly there are an $SU(2)$ worth of gauge transformations which change $\nu_{v_0}$
but otherwise leave $\Eb , s$ invariant. It then follows from Reference \cite{area} that the averaging over
this set of $SU(2)$ transformations with respect to the Haar measure yields a vanishing result.  
Further,  it is easy to see that among these transformations, there are gauge transformations
which rephase $\nu_{v_0}$ by any desired amount. From Reference \cite{area}, such transformations
rephase the state $|s, \Eb \rangle$ by any desired amount so that, our general arguments (\ref{phprop1}), 
once again indicate the vanishing of its group average. 
Moreover our general argument (\ref{rephaseo}) shows that such states lie in a distinct superselection sector
from states which do not admit  rephasings.
Our argumentation then suggests that $\mathring{V}_0$ cannot contain any gauge invariant vertices if the resulting
state is to be non-trivially averaged. 
Note that our argumentation concerning rephasing transformations 
applies equally well to a {\em sum} over gauge related images (as opposed to a Haar measure type integration) if we
agree, as in section III, that an infinite sum over phases vanishes.

Next, consider the rank 1 case so that $V_1$ is of non- zero measure.
Similar to section IIIB let  $\Eb^a = {\hat n} X^a $ 
and consider its symmetries $e^{\theta {\hat n}}$ with $\theta$ supported in $V_1$. Consider a gauge variant vertex 
$v$ of $s$,  $v \in \mathring{V}_1$. It is then easy to check that the integral over $\theta$ of the action of 
these gauge transformations on $\nu_v$ vanishes only if $j_v$ is half integer. 
If $j_v$ is integer valued the integration projects $\nu_v$ into the zero eigenvalue subspace of ${\hat n}$.
Further work is needed to construct the group averaging map to accomodate this result.
Note that in the case of integer $j_v$, Sahlmann's arguments of 
ill definedness of a `sum' hold and one is obliged to use an integral over $\theta$.
In summary, once again, the rank 1 case exhibits subtelities not encountered in the rank 0,2,3 cases.

Let us then restrict attention to the case studied in section \ref{sec7} i.e. the case of $V_1$ of measure
zero. In this case
we take our rough arguments above as indicative that KS states with gauge variant vertices in $\mathring{V_{0}}$
consistently average to zero. Thus all gauge variant vertices must be confined to ${\bar V}_2$.
Let us then restrict attention to the (putative) superselection sector of states with generic $\Eb$ in $V_2$
and $s$ such that its  gauge variant vertices lie only in ${\bar V}_2$.
Since for generic $\Eb$, the only elements of $\aut$ which preserve $\Eb$ are identity on ${\bar V}_2$,
there seems to be no obstruction to using  a sum over states form of the averaging map applied to $|s, \Eb\ket$
and we anticipate that with this minor generalization, all the considerations of \ref{sec7} can be repeated without
obstruction.

\subsection{Summary of technical assumptions.}\label{sec9C}
In the main body of the paper, we have made certain technical assumptions. We list them and discuss their
status below:

\noindent (1) We assume that semianalytic $C^k$ vector fields on $\Sigma$ generate semianalytic $C^{k+1}$ diffeomorphisms. 
As far as we can 
tell this is implicitly assumed by researchers in LQG but we have never seen a proof of the statement.
Proofs are available for the smooth \cite{amp} and analytic categories (see for example \cite{menewrep}) and we think it is a reasonable assumption to make for the semianalytic category as well.
\\

\noindent (2) We have assumed that the quotient groups, $D_{\psi}$ of section \ref{sec7}, are discrete and finite.
We think that this a plausible assumption modulo point (3) below.
Nevertheless, even if  the assumption is false, it does not necessarily
mean that the group averaging map does not exist. Instead, we anticipate that if these groups are either
continuous or discrete but infinite,  there will be further superselection. More work would then be needed to 
see if the ambiguity parameters in these superselection sectors can be  consistently determined.

Note that we have been careful to define the `reference' group $\sym^0_{\psi}$ of equation (\ref{sym0KS02})
in terms of automorphisms which are identity on ${\bar V}_2$ {\em irrespective of their extension into} $V_0$.
This is to avoid  potential `infinite discreteness from topology'  of the type  encountered 
in section \ref{sec8C} in the rank 1 support set context (see footnote \ref{fncs}). 
\\

\noindent (3) We have provided a `counting'  argument for the genericity in the rank 2 case in Appendix \ref{r2sym}. 
We feel the argument
is a bit formal but that  with further work, it should be possible to be converted into a rigorous definition of genericity.
\\

\noindent (4) We have assumed the existence of observables which map the finite span of spinnets into itself.
This assumption also underlies the LQG analysis (see for example \cite{almmt} for the real analytic category).
While such  operators certainly do exist at the quantum level as linear maps from the finite span of 
spinnets into itself, it is not clear to us if classical correspondents to these observables exist. 
Indeed in the case of LQG, apart from the operator 
corresponding to the volume of the entire spatial slice, we do not know of {\em any} such operators which arise
from the {\em quantization} of {\em classical}
observables. A set of classical observables which are gauge and diffeomorphism invariant 
correspond to integrals of density weight one integrands constructed exclusively from phase space variables.
Such integrands can be constructed from the triad and curvature and, in general, are expected to 
be connection dependent and may not even
admit quantum analogs on the diffeomorphism (or automorphism) invariant Hilbert space which arises from group averaging.
On the other hand, it may be the case, that the natural observables relevant to the interpretation of 
quantum geometry are not of this type.
% and should perhaps be thought of as `evolving constants of motion' \cite{rovelli}.
In any case, at the very least, we feel that the group averaged Hilbert spaces constructed here and in LQG
(or their `habitat' deformation \cite{habitat}) serve as useful arenas to explore properties of the Hamiltonian
constraint \cite{habitat,meanomalyfree,alokanomalyfree}.

\subsection{Avenues for further work} \label{sec9D}

For the case of  rank 1 support set 
of non- zero measure, 
much more work is needed
to establish the existence of superselection sectors in which the group averaging map is well defined
and unambiguous. As a first step, it would be very useful to characterise the conditions under which 
background  fields with rank 1 sets of non- zero measure admit symmetries generated by vector fields
transverse to the direction of the field (see end of section \ref{sec8A} and appendix \ref{symr1}). We anticipate
that this would go a long way to supplying ingredient (i) of section \ref{sec9A} above.

Let us  restrict attention to the well understood (modulo assumption (3) of section \ref{sec9C})
 superselection sector of general KS states 
with backgrounds whose rank 1 set is of zero measure. Then, 
the most direct application of our work here is, as mentioned in section \ref{sec1}, in the context of asymptotically flat space
times. While our work here is in the context of compact manifolds, the spatial slice in the asymptotically flat context 
are non- compact. However, the topology of such slices is fixed outside a compact region to be that of the complement of
a compact region in 
$\reals^3$ (see for example \cite{aabook}). In this $\reals^3$ region, the triad boundary conditions imply, in particular,
that the triad is non-degenerate and hence of fixed rank 3. Moreover, the automorphism group consists of those
automorphisms which approach identity at spatial infinity. We expect that these facts should allow
us to identify an Aut invariant superselection sector  which is obtained by the group averaging of KS spinnets
for which (a) the spinnet graph is confined to a compact set  where the background vanishes, (b)
the background triad label asymptotes
to the fixed flat triad at spatial infinity and (c) the triad label is exclusively of rank 0 or 2,3 almost everywhere.
 Work on this is in progress \cite{memiguelaf}.  Of course, the Hamiltonian constraint may map states  out of the Aut invariant superselection sector 
under consideration. Moreover, as mentioned in section \ref{sec1}, the use of the  KS representation is only a first step;
we expect the final picture to involve only the LQG sector, suitably generalised to admit states for which 
the asymptotically flat boundary conditions are satisfied in a suitably coarse grained sense.
However, since very little is known about asymptotically flat kinematics,
let alone the situation at the Aut invariant level, we believe that it is still a useful exercise to 
enquire as to how
asymptotically flat boundary conditions together with Aut invariance can be imposed in the KS
representation in the limited manner envisaged above.

As sketched above the KS representation can be used to satisfy the asymptotically flat boundary conditions on the 
triad. However, there are also the conditions on the connection. We believe that the incorporation of these
conditions on the connection is a more subtle matter and should be built into  the very construction of 
the representation. More in detail, work in progress seems to show, similar to the LQG case \cite{abar},
 that the KS representation in the compact case
can be understood as an $L^2$ representation on a suitable completion of the classical configuration space of connections.
We are hopeful that a generalization of this (putative) result exists for the asymptotically flat case and that we can
interpret the existence of this generalization as the incorporation of the boundary conditions on the connection.

Let us turn briefly from kinematics to dynamics. One of the key open issues in LQG is that of a satisfactory
definition of the Hamiltonian constraint. Recently Laddha  proposed a model with internal gauge group $U(1)^3$ 
as good toy model to formulate such definitions. This model is of interest in its own right
as it corresponds to Smolin's novel weak coupling limit of Euclidean gravity \cite{smolin}. Smolin's idea was that this model
could offer a background independent setting about which one could define Euclidean gravity as a perturbation theory.
While there have been attempts to define the quantum theory of this model in standard LQG like representations
\cite{meanomalyfree,alokanomalyfree}, it is certainly of interest to explore alternate
 background independent quantizations.
Note that the `pure background' sector of the  KS representation provides a quantum 
representation of the algebra of fluxes and background 
exponentials. Since the fluxes and background exponentials do seperate points on phase space, it follows
that this pure background sector does provide an alternative background independent quantization.
It is then of interest to enquire if the pure background representation 
for internal gauge group $U(1)^3$ together with
its Aut averaging provides an alternate setting in which to explore the treatment of the Hamiltonian constraint in the model.  Finally, another model of interest is that of parametrized field theory (PFT) \cite{pft}.   Recently Sengupta analyzed  the quantization of PFT in a KS-like representation \cite{sandipan1,sandipan2}. This work provides  valuable hints for future studies of LQG dynamics in the KS representation. 

Returning from dynamics to kinematics, we touch on a final issue. The pure background KS representation
and its group averaging are defined for semianalytic background fields. 
As mentioned above, this representation 
can be thought of as arising from a
quantization of the classical algebra of electric fluxes and background exponentials. The background exponentials
are exponentials of the connection integrated against the 3 dimensional, semianalytic background electric field.
What if we replace these 3 dimensional smearings with 1 dimensional ones concentrated around loops in the spatial
manifold? Well, one would then obtain an algebra of electric fluxes and effectively `$U(1)^3$ holonomies'! One could
then enquire as to what would happen if we imposed $SU(2)$ invariance in the resulting KS representation.
A heuristic treatment is outlined in Appendix \ref{wilsonloop}, the main result being a hint that what one obtains
is  the standard LQG $SU(2)$ representation in terms of $U(1)^3$ structures. It would be of interest to see
if this treatment can be made less heuristic and if such a less- heuristic treatment leads to useful
insights or applications in the standard LQG context.

%\section*{Acknowledgments}
\textbf{Acknowledgements:} 
We would like to thank Alok Laddha, Hanno Sahlmann, Joseph Samuel and Sandipan Sengupta for helpful discussions. We are very grateful to Hanno Sahlmann for his constant encouragement.

\appendix

\section{Phase factors}\label{appph}

\subsection{Gauge invariance of phase factors associated to symmetries} \label{ginvalpha}
Let $\Eb$ be any triad, and   $b \in \S_{\Eb}$ an element of its symmetry group. Let $a$ be any element in $\aut$ and define $\Eb' := a \cdot \Eb $ and  $b'  :=  a b  a^{-1}$
so that $b' \in \S_{\Eb'}$.  We  show that the phase is gauge invariant in the sense that $e^{i \alpha(b',\Eb')}=e^{i \alpha(b,\Eb)}$.   First notice that the phase associated to the symmetry $b$ of the triad $\Eb$ can be written as
\be
e^{i \alpha(b,\Eb)}= \beta^*_{\Eb} \; (b \cdot \beta_{\Eb}) , \quad \text{for} \quad b \in \S_{\Eb},
\ee
where $\beta^*_{\Eb}$ is the complex conjugated of $\beta_{\Eb}$ so that $\beta_{\Eb}^*\beta_{\Eb}=1$. Thus
\ba
e^{i \alpha(b',\Eb')} & = & \beta^*_{\Eb'} \; (b' \cdot \beta_{\Eb'}) \\
& = & (a \cdot \beta_{\Eb})^* (a b a^{-1}) \cdot (a \cdot \beta_{\Eb})\\
& = & e^{i \alpha(b,\Eb)} . \label{eas}
%& = &(a \cdot \beta_{\Eb})^* (a \cdot \beta_{\Eb})
\ea
A consequence of this result is that the existence of nontrivial rephasings is a gauge invariant property in the sense that  $\sym_{\Eb} \subsetneq \S_{\Eb} \implies \sym_{a \cdot \Eb} \subsetneq \S_{a \cdot \Eb}$.
%Finally, if we take $s$ to be a 1 parameter subgroup of symmetries, from the previous section we know its phase is of the form $\alpha(t)=t k$ where $k=\dot{\alpha}$ is a constant given by (\ref{alphadot}). Thus  (\ref{eas}) translates into $e^{i t k}=e^{i t k'} \; \forall t$ so that  $k=k'$ from which we conclude that  $\alpha(s',\Eb')=\alpha(s,\Eb)$.

\subsection{Phase factor for one parameter subgroup of symmetries}\label{phaseat}
Given a triad $\Eb$, an infinitesimal $\aut$ transformation $(\Lambda,\xi) $ satisfying  
\be
(\Lambda,\xi) \cdot \Eb^a \equiv [\Lambda,\Eb^a]- \L_\xi \Eb^a =0, \label{lamxiE}
\ee
generates the one parameter subgroup of symmetries of $\Eb$:
\be
a_t :=e^{t (\Lambda,\xi) \cdot} =: (g_t,\phi_t).\label{at}
\ee
The condition $\dot{a}_t = (\Lambda,\xi) a_t$ translates into 
\ba
\dot{\phi_t}&=& \L_\xi \phi_t \label{dotphit}\\
\dot{g}_t &=&  \Lambda g_t-\L_\xi g_t \label{dotgt}.
\ea
Thus, whereas $\phi_t$ is simply given by $\phi_t=e^{t \L_\xi}$,  $g_t$ is not in general a 1 parameter subgroup of $\G$. 
Let
\ba
\alpha(t) &:=& \alpha(a_t,  \Eb) \\
&=& \int_{\Sigma} \tr[(\phi_t)_* (\Eb^a) g_t^{-1}\partial_a g_t].
\ea
Using (\ref{dotphit}), (\ref{dotgt}), and the fact that $g_t (\phi_t)_*(\Eb^a)g_t^{-1}=\Eb^a$,  a straightforward computation leads to:
\be
\dot{\alpha}(t)=\int_{\Sigma} \tr[ \Eb^a \partial_a \Lambda]. \label{alphadot}
\ee
Thus $\dot{\alpha}(t)$ is independent of $t$. Since $\alpha(0)=0$ we conclude
\be
\alpha(t)=t \int_{\Sigma} \tr[ \Eb^a \partial_a \Lambda]. \label{alphat}
\ee
Finally, setting $t=1$ we obtain Eq. (\ref{formulalpha}).

\subsection{The case of rank 3 triads} \label{commentr3}
It was noted in section \ref{sec4B} that  phases  associated to  symmetric rank 3 configurations are necessarily zero if the symmetry group does not admit non-trivial homorphisms to $U(1)$.  Here we identify  additional conditions  in which we can guarantee the phase is vanishing. 

Consider  a case where the metric associated to the rank 3 triad $\Eb^a$ admits a Killing vector field $\xi^a$, i.e.:
\be
\tr[ \Eb^{(a} \L_\xi \Eb^{b)}]=0. \label{lieEE}
\ee
Using the $su(2)$ matrices identity $[[a,b],c]= \tr[a c] b-\tr[b c]a$ one can verify that $(\Lambda,\xi)$ is a symmetry of $\Eb^a$ if $\Lambda$ is chosen as:
\be
\Lambda = \tfrac{1}{2}[\Eb_a , \L_\xi \Eb^a].
\ee
Let $\Eb^a=\Eb^a_i \t_i$. If there exist a gauge choice such that 
\be
\Eb^a_3 \propto \xi^a,\label{gauger3}
\ee
 then the action of $\xi^a$ is locally a rotation in the $(\Eb^a_1,\Eb^a_2)$ plane, and thus  $\Lambda$ is proportional to $\tau_3$. From (\ref{formulalpha}) we conclude that $\alpha(e^{(\Lambda,\xi)\cdot},\Eb)=  \int \partial_a \Eb^a_3 \Lambda_3$. But since $\Eb^a_3$ is parallel to $\xi^a$, and $\xi^a$ is divergence free, it follows that $\partial_a \Eb^a_3=0$. Thus if the gauge choice (\ref{gauger3}) is available, we are guaranteed the  phases associated to $\xi^a$ are zero.

\section{Results on semianalytic category} \label{saapp}

We first recall some definitions and results from Appendix A of \cite{lost}. In order to define semianalytic manifolds and functions, one first  defines   semianalytic functions on  open sets of $\reals^n$.  That  requires the notion of \emph{semianalytic partition} of an open set $U \subset \reals^n$. This is given by a decomposition of the form:
\be
U=\cup_{\sigma} V_\sigma,  \quad V_\sigma \cap V_{\sigma'} =\emptyset \; \text{if } \; \sigma \neq \sigma' ,\label{saf1}
\ee
where  $V_\sigma$ are described by analytic functions  $h_{i}, i=1,\ldots n$, on $U$ according to: 
\be 
V_\sigma \equiv V_{\sigma_1, \ldots \sigma_n}:= \cap_{i=1}^n \{h_i(x) \sigma_i 0 \} , \quad \sigma_i =\{ 0, > ,< \}. \label{saf2}
\ee
A key property of semianalytic partitions (see proposition A.9 of \cite{lost}) is that  every $x \in U$ has a neighborhood $U_x$ such that  $V_\sigma \cap U_x \subset U_x$ is a finite union of analytic submanifolds of $U_x$. This  result  will allow us to conclude the sets  (\ref{V0}),(\ref{V1}) and (\ref{V2}) are finite unions of submanifolds. 

Given an open set $U \subset \reals^n$, a function $f:U  \to \reals$ is said to be semianalytic if for every $x \in U$ there exists an open neighborhood $U_x$ with a semianalyitic partition as above  such that 
\be
f|_{V_\sigma}=f_\sigma|_{V_\sigma} ,\label{saf3}
\ee
 where $f_{\sigma}$ are analytic functions on $U_x$. 
 
 Semianalytic manifolds and submanifolds are then defined  as in the  differentiable case, with the  requirement that  transition functions between local charts are semianalytic \cite{lost}. Likewise semianalytic functions on such manifolds are defined by the requirement that they are semianalytic on the local charts. 

The main property of semianalytic functions that we will need is  the following:

\emph{Given semianalytic functions $f_i : \Sigma \to \reals, i=1,\ldots n$ and a choice $\sigma_i = <, >,0$ for each $i$,  the set $X=\cap_{i=1}^n \{ f_i \sigma_i 0 \} \subset \Sigma$ can be decomposed into a  finite union of submanifolds.} 

For the sake of clarity,  we show this result for the case of a single function,  $X:=\{f(x)= 0\}$. From the structure of the proof it will become evident that the general case follows.  Let $\{ \chi_I, U_I \}$ be a semianalytic atlas of $\Sigma$ compatible with $f$. That is,  there is a semianalytic partition on each local chart:
\be
U'_I:=\chi_I(U_I)=\cup_{\sigma} V^I_\sigma, 
\ee
as in (\ref{saf1}), (\ref{saf2}) such that,
\be
f \circ \chi^{-1}_I |_{V^I_\sigma}=f^I_\sigma|_{V^I_\sigma} ,
\ee
where $f^I_\sigma$ are analytic functions on  $U'_I \subset \reals^3$.  On each local chart the set of interest takes the form:
 \be
\chi_I(X \cap U_I)= \cup_{\sigma} \{f^I_{\sigma}(x)=0 \} \cap\{ h^I_1(x) \sigma_1 0 \} \ldots \cap  \{ h^I_{n_I}(x) \sigma_{n_I} 0 \}. \label{zls}
 \ee
The sets featuring in the union (\ref{zls}) can be realized as sets of a new  partition of $U'_I$ defined in terms of the functions $\{h^I_i \} \cup \{ f^I_{\sigma} \}$. By the proposition of \cite{lost} mentioned  after Eq. (\ref{saf3}), it follows that  every $x \in U_I$ has  an open neighborhood $U^x_I$ such that
\be
\chi_I(X \cap U^x_I) =  \text{finite union of analaytic submanifolds}.
\ee
This in turn implies that $X \cap U^x_I$ is a finite union of \emph{semianalytic} submanifolds of $\Sigma$. Since $\Sigma$ is compact, there exist a finite subcover of the (uncountable) open cover   $ \{ U^x_I \}$. This allows to express $X$ as a finite union of submanifolds.% @@@@say few words about how non-compact $\Sigma$@@@@

\section{Additional proofs}\label{random}
\subsection{Generic rank 2  $\tilde{\tilde{q}}^{ab}$ do not admit symmetries}\label{r2sym}
Rank 2 symmetric tensors $\tilde{\tilde{q}}^{ab}$ are characterized by the existence of a degenerate  direction:
\be
\exists \; \lambda_a \; : \; \tilde{\tilde{q}}^{ab}\lambda_b \; = \; 0. \label{qlambda}
\ee
The number of independent components that are needed to specify such tensors can be counted as follows. A general symmetric tensor has 6 independent components. We then need to specify the degenerate direction $\lambda_a$, which  requires 2 independent components. Finally condition (\ref{qlambda}) gives 3 constraints on the components of the tensor. Thus the number of independent functions to define such tensor is given by 6+2-3=5.  This in turn implies that the equation 
\be
\L_{\xi}\tilde{\tilde{q}}^{ab}=0 \label{lieqtt2}
\ee 
will generically give 5 independent conditions for the vector field $\xi^a$ (which represents  3 unknowns to be solved for). The system   is thus generically overdetermined and hence there are no  generic solutions to (\ref{lieqtt2}).

\subsection{Local description of rank 1 configurations and their symmetries} \label{symr1}
Consider a triad $\Eb^a$ with a non-trivial rank 1 region in the sense that there exist an open set $U \subset \Sigma$  where $\rk(\Eb)|_U=1$. On $U$ we have the splitting $\Eb^a= \nh X^a$, and we restrict attention to the  case where $X^a$ is divergence free, since otherwise the corresponding state vanishes upon averaging.  An infinitesimal symmetry $(\Lambda,\xi)$ of $\Eb$ will then have to satisfy equations (\ref{lamxi1}) and (\ref{Xxi}) on $U$ (for simplicity we take $U$ to be simply connected). We now describe the local form of the vector fields $\xi^a$ satisfying (\ref{Xxi}). 

As discussed in section \ref{sec8A}, the divergence free property of $X^a$ translates in the closeness of the two form $\w_{ab}=\eta_{abc}X^c$. This means that  $\w_{ab}$ is a  presymplectic form on $U$. Darboux theorem for presymplectic forms (see for instance theorem 5.1.3 of \cite{abmad})  tell us that every point in $U$ has an open neighborhood with local coordinates  $x,y,z$ such that:
\ba
\w_{ab} & = & dy \wedge dz  \label{localw} \\
X^a & = & \frac{\partial}{\partial x}. \label{localX}
\ea
%(We emphasize that  (\ref{localw}) and (\ref{localX}) in principle  hold for sufficiently small regions of $U$). 
It is easy to see that the most general solution to $\L_{\xi}X^a=0$ in the local coordinates takes the form:
\be
\xi^a=  g(x,y,z) \frac{\partial}{\partial x}+ \partial_y f(y,z) \frac{\partial}{\partial z}- \partial_z f(y,z) \frac{\partial}{\partial y} \label{vflc} 
\ee
for arbitrary functions $f(y,z)$ and $g(x,y,z)$. In the language of section \ref{sec8A},   the first term correspond to  vector fields parallel to $X^a$ and the  remaining part corresponds to the   `transversal' vector fields. Notice that this transversal part  takes the form of a `Hamiltonian vector field', with  `Hamiltonian' $f$ and symplectic form (\ref{localw}). The function $f$ in (\ref{vflc}) corresponds to the  function $f$ in (\ref{Xxi}), and  the independence on the $x$ variable corresponds to the condition  $X^a \partial_a f=0$ that follows from (\ref{Xxi}). This condition  forbids a choice of  $f$ with a compact support inside the Darboux chart. Thus, to obtain a well-defined vector field $\xi^a$ associated to a non-trivial $f$ one has to appropriately patch together the solutions (\ref{vflc}) for the different Darboux charts. Furthermore, in the varying rank case, one would  need to take into account  the regions where the rank changes from 1 to $\neq 1$. At such points  Darboux charts  are no longer available and one would need new local descriptions. Whereas it is clear how one should proceed for a given particular configuration, we are not able to give further characterization of these `transversal' symmetries in a generic setting. Such characterization is also needed to determine the phases of the varying rank configurations in the case where  $\partial \Vo_1$ is not connected, see Eq. (\ref{finalalpha}).

\subsection{Eq. (\ref{cond3})} \label{secC4}
Let $\eo$ denote the open edge that results from removing the endpoints of a closed edge $e$. Similarly, given a graph  $\gamma=\cup_{n=1}^{N}e_n$ composed of $N$ closed edges, we define  $\go:=\cup_{n=1}^{N}\eo_n$. The following lemma will be used in obtaining (\ref{cond3}): 

\emph{Given a semianalytic graph $\gamma$, and a point $p \in \eo \subset \go$, there exists an open neighborhood $U_p$ such that $U_p \cap \go = U_p \cap \eo$.} 

To show this assertion, suppose by contradiction there exists a point $p \in \eo$ that does not admit any such open neighborhood.  Then  $(\go \setminus \eo) \cap U_p \neq \emptyset$ for every open  neighborhood $U_p$ of $p$. It follows that there exists a sequence of points $p_n \in (\go \setminus \eo)$ that have $p$ as an accumulation point. Since $p, p_n \in \gamma$ and $\gamma$ is compact, there exists a subsequence of $p_n$ that converges to $p$. This can only happen if $p$ is a vertex of $\gamma$, which then contradicts the condition $p \in \eo$.

We now show Eq. (\ref{cond3}). Suppose by contradiction that  $s$ and $s'$ satisfy
\be
\gamma(s) \cap \Vo_0 \neq \gamma(s') \cap \Vo_0. \label{cond3app}
\ee
 Let $\gamma (s,s')$ be the coarsest graph  underlying  $s$ and $s'$ and consider the graph $\tg:=\gamma(s,s') \setminus (\gamma(s) \cap \gamma(s'))$. Semianaliticity of the original graphs ensures  $\tg$ is itself  semianalytic  and condition  (\ref{cond3app}) is equivalent to $\tg \cap \Vo_0 \neq \emptyset$.  Since the  intersection of a graph with an open set cannot consist of isolated points (otherwise we would be able to find an open set $U$ whose intersection with the graph consists of a single point),  it follows that there exists an open interval $I \subset \tgo \cap \Vo_0$. $I$ is either contained in  $\gamma(s)$ or in $\gamma(s')$. Consider the case where $I \subset \gamma(s)$. From the lemma above, every point $p \in I$ has an open neighborhood $U_p$  such that $U_p \cap \tg  =U_p \cap I$. Furthermore, we can take $U_p$  such that $U_p \subset \Vo_0$. We can then construct vector fields with support in $U_p$ of the type used in the argument  following Eq. (\ref{strobsarg}) in order to produce  infinitely many diffeomorphisms that change $I$ (and hence $s'$) but leave $s$ unchanged. This implies  $\bra s, \Eb | O |s',  \Eb' \ket$ vanishes, and hence the contradiction.

 \subsection{Eq. (\ref{cond3p})} \label{appcond3p} 

Diffeomorphisms generated by vector fields of the form $\xi^a = g X^{a}$ are symmetries of $X^a$ for any density weight minus one scalar $g$. These in turn generate continuous symmetries of $\Eb$. If $\bra \Eb| O |\Eb' \ket \neq 0$ they must also be symmetries of  $\Eb'$ :
\be
\bra \Eb| O |\Eb' \ket \neq 0  \implies  \L_{g X} X'^a|_{\Vo_{01}} =0 \quad \forall \, \; \text{ density weight   $-1$} \;  g  \; {\rm in} \; \Vo_{01}   \label{symxxp}
\ee
(recall that $\Vo_{01}=\Vo'_{01}$). Let us now use an auxiliary metric $\qo_{ab}$ on $\Sigma$ to express (\ref{symxxp}) in terms of the associated covariant derivative $\Do$, $\Do_c \qo_{ab}=0$: 
\be
\L_{g X} X'^a= g (X^b \Do_b X'^a - X'^b \Do_b X^a) + 2 \Do_b g X^{[b}X'^{a]}=0  \quad \forall g. \label{symxxpq}
\ee
%:
Taking $g= \qo^{-1/2}$, the second term vanishes and we conclude that $(X^b \Do_b X'^a - X'^b \Do_b X^a) =0$. Equation (\ref{symxxpq}) then becomes:
\be
\Do_b g X^{[b}X'^{a]}=0 \quad \forall g  \implies X^{[b}X'^{a]}=0,
\ee
and we recover condition (\ref{cond3}).

\section{Semidirect product structure of  $\aut$}\label{appsd}
Let us first recall the notion of  semi direct product (we follow presentation from \cite{romano}).
Let $D$ and $G$ be two Lie groups and let $\sigma: D \to Aut(G)$ be an homomorphism, that is, for each $\phi \in D$, $\sigma_\phi: G \to G$ is an invertible map satisfying:
\be
\sigma_\phi( g g')=\sigma_\phi(g) \sigma_\phi(g') , \quad \sigma_{\phi \phi'}(g)=\sigma_\phi (\sigma_\phi'(g)), \quad \forall \, g,g' \in G, \phi,\phi' \in D .\label{propsigma}
\ee
The semidirect product $G \rtimes D$ is a new Lie group whose elements are pairs $(g,\phi) \in G \times D$ satisfying the following product:
\be
(g,\phi) (g',\phi'):=(g \sigma_\phi(g'), \phi \phi'). \label{sdprod}
\ee
Properties (\ref{propsigma}) guarantee all the group product rules are satisfied. The Lie algebra structure is as follows.  As a vector space is given by $\Lie(G \rtimes D)=\Lie(G)\oplus \Lie(D)$. The Lie bracket can be obtained from (\ref{sdprod}) and is given by,
\be
[(\Lambda,\xi),(\Lambda',\xi')]=(\tau_\xi(\Lambda')-\tau_{\xi'}(\Lambda)+[\Lambda,\Lambda'],[\xi,\xi'] ) \label{sdlb}
\ee
where $\tau: \Lie(D) \to \Lie(G)$ is a Lie algebra homomorphism induced by $\sigma$ and defined as follows. Let $g(s)$ and $\phi(t)$ be 1-parameter family of group elements such that $g(0)=\id_G, \dot{g}(0)=\Lambda \in \Lie(G)$ and $\phi(0)=\id_D, \dot{\phi}(0)=\xi \in \Lie(D)$. Then
\be
\tau_\xi(\Lambda):= \frac{d}{dt}\frac{d}{ds}\sigma_{\phi(t)}(g(s))|_{t=s=0}.
\ee

In our case of interest we have: $D=\diff$, $G=\G$,  $\sigma_\phi(g)=\phi_* g$ and $\tau_\xi(\Lambda)=-\L_\xi \Lambda $ .

\section{Heuristic relation between spin networks and distributional background triads} \label{wilsonloop}
In an Abelian $U(1)$ gauge theory, holonomies can be realized as background exponentials functions associated to   distributional  electric fields: Given a closed curve $\gamma: [0,1] \to \Sigma$ and $j \in \Z$,  define the distributional electric field
\be
\Eb_{(\gamma,j)}^a(x) :=  j \int dt \, \dot{\gamma}^a(t) \delta(x,\gamma(t))  \label{distX}
\ee
so that
\be
\beta_{\Eb_{(\gamma,j)}}[A] = e^{i \int_{\Sigma} \Eb_{(\gamma,j)}^a A_a} = e^{i j \int dt \dot{\gamma}^a A_a} = h^j_\gamma[A] ,
\ee
where $h^j_\gamma[A] $ is the holonomy along the closed curve $\gamma$ in the $j$ representation of $U(1)$. Notice that $\Eb_{(\gamma,j)}^a$ defined by (\ref{distX}) satisfies the divergence free condition $\partial_a \Eb_{(\gamma,j)}^a=0$. 

The analogue of   (\ref{distX}) in the $SU(2)$ theory is given by a rank 1 triad $\Eb^a = \nh X^a$ with $X^a$ a distributional vector field  with support on  $\gamma$, and $\nh$ a normalized internal direction at each point of  $\gamma$:
\be
\Eb_{(\gamma,j,\nh)}^a(x) :=  j \int dt \, \dot{\gamma}^a(t) \delta(x,\gamma(t))\, \nh(t) , \quad j \in \Z/2 \label{distE},
\ee
so that
\be
\beta_{\Eb_{(\gamma,j,\nh)}}[A]  = e^{i j \int dt \dot{\gamma}^a \tr[ A_a \nh]}  \equiv e^{i j \int_{\gamma} \tr[ A \nh]}, \label{distbe}
\ee
where the last equation represents the integral of the one form $\tr[ A \nh]$ along the 1-dimensional curve defined by  $\gamma(t)$ (which we are assuming has no self-intersections; later we comment on the general graph case). The reason for taking   $j \in \Z/2$ will become clear below. 
Let us now study the behavior of the distributional background exponential (\ref{distbe}) under local $SU(2)$ rotations.  Given $g \in \G$, the phase factor associated to (\ref{distE}) can be expressed as a difference of phases associated to $\nh$ and $g \nh g^{-1}$:
\ba
\alpha(g,{\Eb_{(\gamma,j,\nh)}}) &=& j \int_\gamma \tr[g^{-1}d g \nh]\\
&=& j(\Phi[g \nh g^{-1}]- \Phi[\nh]) \label{phasediff}
\ea
where
\be
\Phi[\nh]= \int_{D}\tr[\nh d \nh \wedge d \nh]
\ee
is  given by a  two dimensional integral over a surface $D$ such that $\gamma = \partial D$.\footnote{If $\gamma$ cannot be realized as the boundary of a two dimensional surface in $\Sigma$, it is still possible to  define $\Phi$ as a two dimensional integral on the parameter space where $t$ belongs.} Geometrically it corresponds to the  area enclosed by   $\nh(t) \subset S^2$ . For $j \in \Z/2$, the phase $e^{i j \Phi[\nh]}$ is independent of whether one takes the `inside' or `outside' area of this  curve.

From (\ref{phasediff}) it follows that it is possible to absorb the occurrence of phases in the definition of the  distributional background exponentials. Let
\ba
w^j_{\gamma, \nh}[A] &:= &e^{i j \Phi[\nh]}\beta_{\Eb_{(\gamma,j,\nh)}}[A], \label{wjgamma} \\
&=& e^{i j \Phi[\nh]}e^{i j \int dt \tr[\nh A_a \dot{\gamma}^a]},
\ea
be the rephased background exponential. It then follows that $w^j_{\gamma, \nh}$ transforms covariantly under the action of $g \in \G$:
\ba
g \cdot w^j_{\gamma, \nh}[A] & \equiv & w^j_{\gamma, \nh}[g^{-1} \cdot A] \\
&=& w^j_{\gamma, g \nh g^{-1}}[A].
\ea
Thus, if we denote by  $|\nh \, \gamma \, j \ket$ the state associated to $w^j_{\gamma, \nh}$, the $\G$- group averaging formula would take the simple form:
\be
\eta(| \nh \, \gamma \, j\ket) = \sum_{\nh(t)} \bra  \nh \, \gamma \, j | , \label{gavfakehol}
\ee 
with the  sum ranging  over all possible internal directions $\nh(t)$ along the curve $\gamma(t)$ and with no additional phases present. 
At a formal level, this expression defines the following `wave function' of the connection $A$:
\be
`` \quad \overline{\eta(| \nh \gamma j \ket) | A \ket}= \sum_{\nh(t)} w^j_{\gamma, \nh}[A]\quad ". \label{gawf}
\ee

It turns out that the formal expression on the right hand side of  (\ref{gawf}) can be identified with a Wilson loop function as follows \cite{diakonov}. Let $h^j_\gamma[A]$ be the holonomy in the $j$ representation along  $\gamma$ and  $W_\gamma^j[A]=\text{trace}(h^j_\gamma[A])$. $h^j_\gamma[A]$ can  formally be thought of as an ``evolution operator'' (in  the spin $j$ vector space) associated with the ``time dependent Hamiltonian'' $H(t):=A^i_a(\gamma(t)) \dot{\gamma}^a(t) \pi^j(\t_i)$, where $\pi^j(\t_i)$ are the $su(2)$ generators in the $j$ representation. One can then express this ``evolution operator'' in a path integral form in the spin $j$ coherent state basis $| \nh \ket$, $\nh \in S^2$. Doing so from $t=0$ to $t=1$ where $\gamma(0)=\gamma(1)$,  and taking the trace, one arrives at  \cite{diakonov}:
\be
W_\gamma^j[A] = \int \D \nh(t) \, w^j_{\gamma, \nh}[A], \label{piwl}
\ee
where  $\int \D \nh(t)= \lim_{N \to \infty} (\tfrac{2 j+1}{4 \pi})^{N}\int d \nh_1 \ldots \int d \nh_{N} $, with  $\int d \nh_n$ the integral on the unit sphere corresponding to the discretized `time' $t_n=n/N$, and   $w^j_{\gamma, \nh}[A]$ given by (\ref{wjgamma}). Expression (\ref{piwl}) has structurally the same form as the formal `group averaged wave function' (\ref{gawf}).

 The above formal procedure can be extended to  general graphs to obtain `spin network wave functions'. In this case, there is additional gauge invariant information encoded in the relative directions of the internal vectors  $\nh_e$ at the vertices where the edges $e$ meet. Upon averaging one obtains a  spin network function in a  coherent intertwiner representation \cite{ls}, with intertwiners  determined by the aforementioned gauge invariant information.

\end{document}